\documentclass[twocolumn]{pasj01}

\Received{$\langle$reception date$\rangle$}
\Accepted{$\langle$acception date$\rangle$}
\Published{$\langle$publication date$\rangle$}

\usepackage[switch, mathlines]{lineno}

\usepackage{lscape}
\usepackage{colortbl}

\begin{document}

\title{Multi-epoch X-ray spectral analysis of Centaurus~A: revealing new constraints on iron emission line origins}
\author{Toshiya Iwata,\altaffilmark{1,}$^{*}$
Atsushi Tanimoto, \altaffilmark{2}
Hirokazu Odaka, \altaffilmark{3,4,6}
Aya Bamba, \altaffilmark{1,3,5}
Yoshiyuki Inoue, \altaffilmark{6,7,4}
and Kouichi Hagino\altaffilmark{1}}%

\altaffiltext{1}{Department of Physics, The University of Tokyo, 7-3-1 Hongo, Bunkyo-ku, Tokyo 113-0033, Japan}

\altaffiltext{2}{Graduate School of Science and Engineering, Kagoshima University, Kagoshima 890-0065, Japan}

\altaffiltext{3}{Research Center for the Early Universe, School of Science, The University of Tokyo, 7-3-1 Hongo, Bunkyo-ku, Tokyo 113-0033, Japan}

\altaffiltext{4}{Kavli Institute for the Physics and Mathematics of the Universe (WPI), UTIAS, The University of Tokyo, Kashiwa, Chiba 277-8583, Japan}

\altaffiltext{5}{Trans-Scale Quantum Science Institute, The University of Tokyo, Tokyo  113-0033, Japan}

\altaffiltext{6}{Department of Earth and Space Science, Graduate School of Science, Osaka University, 1-1 Machikaneyama, Toyonaka, Osaka 560-0043, Japan}

\altaffiltext{7}{Interdisciplinary Theoretical \& Mathematical Science Program (iTHEMS), RIKEN, 2-1 Hirosawa, Saitama 351-0198, Japan}

\email{toshiya.iwata@phys.s.u-tokyo.ac.jp}

\KeyWords{galaxies: active --- galaxies: individual (Centaurus A) --- X-rays: galaxies}

\maketitle

\begin{abstract}
We conduct X-ray reverberation mapping and spectral analysis of the radio galaxy Centaurus~A to uncover its central structure. We compare the light curve of the hard X-ray continuum from Swift Burst Alert Telescope observations with that of the Fe~K$\alpha$ fluorescence line, derived from the Nuclear Spectroscopic Telescope Array (NuSTAR), Suzaku, XMM-Newton, and Swift X-ray Telescope observations. The analysis of the light curves suggests that a top-hat transfer function, commonly employed in reverberation mapping studies, is improbable. Instead, the relation between these light curves can be described by a transfer function featuring two components: one with a lag of $0.19_{- 0.02}^{+ 0.10}~\mathrm{pc}/c$, and another originating at $r > 1.7~\mathrm{pc}$ that produces an almost constant light curve.
Further, we analyze the four-epoch NuSTAR and six-epoch Suzaku spectra, considering the time lag of the reflection component relative to the primary continuum. This spectral analysis supports that the reflecting material is Compton-thin, with $N_{\mathrm{H}} = 3.14_{-0.74}^{+0.44} \times 10^{23}~ \mathrm{cm}^{-2}$. These results suggest that the Fe~K$\alpha$ emission may originate from Compton-thin circumnuclear material located at sub-parsec scale, likely a dust torus, and materials at a greater distance.
\end{abstract}

\section{Introduction}
\label{intro}

Centaurus A (Cen~A) is one of the closest radio galaxies at 3.8 Mpc (\cite{Harris2010}), hosting a powerful jet powered by a central supermassive black hole (SMBH) with a mass of $5\times 10^7M_{\odot}$ \citep{Neumayer2007}.
Observations of Cen~A span a broad spectrum of energies, from radio to gamma-ray (e.g., \cite{HESS2018}). Similar to Seyfert 2 galaxies, Cen~A is presumed to be viewed through an AGN torus.
Cen~A has been repeatedly observed in the X-ray energy band (e.g., \cite{Evans2004}; \cite{Markowitz2007}; \cite{Fukazawa2011b}; \cite{Furst2016}), making it an excellent subject for studying the circumnuclear environment of central SMBHs in active galactic nuclei (AGNs) with jets. Variations in the absorbing column density support the hypothesis that the torus consists of clumpy materials (\cite{Rothschild2011}; \cite{Rivers2011}).

The X-ray spectrum of Cen~A reveals an iron emission line at approximately $6.4~\mathrm{keV}$ in the rest frame (e.g., \cite{Mushotzky1978}), thought to originate from a reflector irradiated by the central X-ray source. This line is expected to provide insights into the circumnuclear environment, although its exact origin remains uncertain. Analysis of the Fe~K$\alpha$ line profile using the Chandra High Energy Transmission Grating (HETG) indicates it arises from cool, distant material relative to the central SMBH (e.g., \cite{Evans2004}; \cite{Shu2011}).
\citet{Evans2004} analyzed the X-ray spectrum obtained with the Chandra/HETG and estimated the full width at half-maximum (FWHM) velocity ($v_{\mathrm{FWHM}}$) between $1000~\mathrm{km~s^{-1}}$ and $3000~\mathrm{km~s^{-1}}$. Assuming that the relation between $r$ and $v_{\mathrm{FWHM}}$ can be written as $r = 4GM_{\mathrm{BH}}/3v_{\mathrm{FWHM}}^{2}$ (\cite{Netzer1990}; see section 5.1, 1st paragraph, for more detailed information), these values suggest distances of $0.2~\mathrm{pc}$ and $0.03~\mathrm{pc}$. In contrast, the variability of the Fe~K$\alpha$ line flux suggests that it is emitted from material at least a parsec away. \citet{Furst2016} highlighted that the stable Fe~K$\alpha$ line flux (e.g., \cite{Rothschild2006}; \cite{Rothschild2011}) suggests the emitting region is located at least 10 lt-yr (approximately $ 3~\mathrm{pc}$) or more away from the core.

X-ray reverberation mapping (\cite{Uttley2014}) using the Fe~K$\alpha$ line (e.g., \cite{Ponti2013}; \cite{Zoghbi2019}; \cite{Andonie2022}) is poised to to further constrain its origin. The flux variability of the Fe~K$\alpha$ line lags behind that of the direct components, attributable to their different light travel distances. Therefore, the location of the reflector can be constrained by comparing their light curves and estimating the time lag between them. This method potentially constrains the size of a parsec or sub-parsec scale reflector based on multi-year observations. 

In addition to the size of the reflector, the optical depth for Compton scattering of the reflector in Cen~A also remains uncertain. The spectral shape of the reflection continuum, produced by Compton scattering at the reflector depends on the Compton thickness of the reflector. When the reflector is Compton-thick, a prominent reflection continuum with a peak at approximately $\sim 10$--$30~\mathrm{keV}$ called ``Compton hump," is expected \citep{Ross2005}. The Compton hump has been observed in various AGNs (e.g., \cite{Risaliti2013}; \cite{Marinucci2014}; \cite{Parker2014}). For Cen~A, some studies have indicated that the Compton-thick reflection model adequately explains the hard X-ray spectra (e.g., \cite{Fukazawa2011b}; \cite{Burke2014}), while others have supported the Compton-thin model (e.g., \cite{Markowitz2007}; \cite{Furst2016}; \cite{Ogawa2021}).

In this paper, we conduct X-ray reverberation mapping and multi-epoch spectral analysis of Cen~A to determine the size and Compton thickness of the reflector. 
For reverberation mapping, we compare the light curve of the hard X-ray continuum from the Neil Gehrels Swift Burst Alert Telescope (Swift/BAT: \cite{Gehrels2004}; \cite{Barthelmy2005}) with that of the Fe~K$\alpha$ fluorescence line from the Nuclear Spectroscopic Telescope Array (NuSTAR: \cite{Harrison2013}), Suzaku \citep{Mitsuda2007}, XMM-Newton \citep{Jansen2001}, and the Swift X-ray Telescope (XRT; \cite{Burrows2005}) observations. Estimating the time lag of the reflection component from this comparison helps ascertain the typical distance and size of the reflector.
Further constraints on the reflector are obtained through multi-epoch spectral analysis using NuSTAR (\cite{Harrison2013}). We adjust the normalization of the reflection component based on the light curve analysis results to account for the time lag of the reflection component. We employ a clumpy torus model (XClumpy: \cite{Tanimoto2019}), accommodating both Compton-thick and Compton-thin cases. Note that \citet{Kang2020} used three out of four NuSTAR datasets but applied a Compton-thick reflection model, which does not consistently explain the Fe~K$\mathrm{\alpha}$ line and the reflection continuum (\cite{Furst2016}).

The paper is organized as follows: section \ref{observation} provides an overview of the data used in this study and the data reduction process. In section \ref{rev_mapping}, we analyze the light curves of the direct component and the Fe~K$\mathrm{\alpha}$ line to derive the time lag of the reflection component. In section \ref{spectral}, we conduct spectral analysis of the four-epoch NuSTAR observations, considering the time lag of the reflection component. We discuss our results in section \ref{discussion} and summarize the key points in section~\ref{conclusion}.

\section{Data reduction}
\label{observation}
We used archived data from NuSTAR, Suzaku, XMM-Newton, Swift/XRT, and Swift/BAT. Details of the data used in this study are shown in table \ref{tab:datalist}.

\subsection{NuSTAR}
 NuSTAR consists of two co-aligned grazing incidence telescopes that focus hard X-rays onto focal plane modules (FPM) A and B consisting of cadmium-zinc-telluride pixel detectors. Cen~A was observed six times with NuSTAR. We analyzed data from four observations (2013, 2015, 2018, and 2019) with exposure times exceeding $10~\mathrm{ks}$. These data were processed using NuSTARDAS, part of HEASoft v6.28, and NuSTAR CALDB version 20210315. Source spectra were extracted from circular regions with a radius of 100 arcsec, and background spectra from source-free circular regions with a radius of 120 arcsec. The NuSTAR spectra were rebinned to ensure each bin contained 50 or more counts.

 \subsection{Suzaku}
 The data from Suzaku were calibrated and screened using the aepipeline within HEASoft v6.28 and Suzaku CALDB XIS 20181010. We extracted the source spectra from annular regions to mitigate the pile-up effect. The inner and outer radii for each dataset were as follows: $45$--$120$~arcsec (100005010), $60$--$240$~arcsec (704018010), $60$--$240$~arcsec (704018020), $75$--$240$~arcsec (704018030), $45$--$240$~arcsec (708036010), and $45$--$240$~arcsec (708036020). The background spectra were extracted from source-free circular regions with a 120 arcsec radius. Suzaku has four sets of X-ray Imaging Spectrometers (XIS: \cite{Koyama2007}). Three XIS CCDs (XIS 0, 2, and 3) are front-illuminated (FI) and the other (XIS 1) is back-illuminated (BI). Spectra from XIS 0, 2, and 3 were combined, and the redistributed matrix files and the auxiliary response files were generated using xisrmfgen and xissimarfgen \citep{Ishisaki2007}, respectively. Since we limited our analysis to the Fe~K$\alpha$ line flux in section \ref{rev_mapping}, the data from Suzaku's Hard X-ray Detector (HXD) were used only in section \ref{spectral}. All spectra were binned to contain at least 50 counts per bin.

\subsection{XMM-Newton}
We processed the XMM-Newton data using the Science Analysis System version \texttt{xmmsas\_20230412\_1735-21.0.0}. Following \citet{Furst2016}, we extracted the source spectra from annular regions with a 10~arcsec inner radius and a 40~arcsec outer radius from the EPIC-pn camera \citep{Struder2001} to address the pile-up effect. Background spectra were taken from a source-free circular region with a 40~arcsec radius. Due to significant pile-up and inadequate photon statistics, we did not use data from the MOS cameras. The spectra were rebinned to ensure a minimum of 50 counts per bin.

\subsection{Swift/XRT}
We analyzed the Swift/XRT data recorded in Windowed Timing mode from February to March and May to July 2012. Data from outside these periods were excluded due to insufficient exposure time and limited observation duration, which was approximately three months. Observations taken in Photon Counting mode were not used due to severe pile-up issues and inadequate photon statistics.

The data were processed using the
XRTDAS software integrated into HEASoft v6.29. The xrtpipeline (version 0.13.6) was used for cleaning and calibrating the event files. Source spectra were extracted from a circular region with an 80 arcsec radius, while background spectra were taken from a source-free annular region with inner and outer radii of 150~arcsec and 300~arcsec, respectively, using the tool xrtproducts (version 0.4.2).

\subsection{Swift/BAT}
The BAT, an instrument on the Swift Observatory, provided light curves sourced from the Swift/BAT hard X-ray transient monitor website \citep{Krimm2013}\footnote{$\langle$https://swift.gsfc.nasa.gov/results/transients/$\rangle$}. These were rebinned to 20-day intervals and the 15--50 keV count rates $\mathrm{CR}_{15-50}$, in units of $\mathrm{counts~cm^{-2} ~s^{-1}}$, were converted to unabsorbed 2--10 keV fluxes using the formula $\mathrm{F}_{2-10}~(10^{-11}~\mathrm{erg~cm^{-2}~s^{-1}})= 5726\cdot\mathrm{CR}_{15-50}$, following \citet{Borkar2021}. The resulting light curve is shown in figure~\ref{lc_sw_fe}.

\begin{table*}

\tbl{Summary of the observational data of Cen~A.}{%
    \begin{tabular}{cccccc}
\hline
\hline
Observatory &ObsID\footnotemark[$*$] & Start\footnotemark[$\dagger$] & End\footnotemark[$\ddagger$] & MJD range &Exposure\footnotemark[$\S$]  \\
\hline
 NuSTAR & 60001081002 & 2013-08-06 & 2013-08-07 &  56510--56511 &  51 \\
 & 60101063002 & 2015-05-17 & 2015-05-18 & 57159--57159  & 23 \\
 & 60466005002 & 2018-04-23 & 2018-04-23 & 58230--58231  & 17 \\
 & 10502008002 & 2019-08-05 & 2019-08-05 & 58699--58700  & 22 \\
\hline
 Suzaku & 100005010 & 2005-08-19 & 2005-08-20 &  53600--53602 &  65 \\
 & 704018010 & 2009-07-20 & 2009-07-21 & 55031--55033  & 62 \\
 & 704018020 & 2009-08-05 & 2009-08-06 & 55047--55049  & 51 \\
 & 704018030 & 2009-08-14 & 2009-08-16 & 55057--55058  & 56 \\
 & 708036010 & 2013-08-15 & 2013-08-15 & 56518--56519  & 11 \\
 & 708036020 & 2014-01-06 & 2014-01-06 & 56663--56663  & 7.4 \\
\hline
 XMM-Newton & 0724060501 & 2013-07-12 & 2013-07-12 &  56485--56485 &  7.3 \\
 & 0724060601 & 2013-08-07 & 2013-08-07 & 56511--56511  & 7.3 \\
 & 0724060701 & 2014-01-06 & 2014-01-07 & 56663--56663  & 17 \\
 & 0724060801 & 2014-02-09 & 2014-02-09 & 56697--56697  & 13 \\
\hline
 Swift/XRT & 00031312009--00031312038 & 2012-02-02 & 2012-03-31 &  55959--56017 &  25 \\
 & 00031312050--00031312094 & 2012-05-02 & 2012-07-31 & 56049--56139  & 43 \\
\hline
\end{tabular}}\label{tab:datalist}
\begin{tabnote}
\footnotemark[$*$]Observation identification string. For Swift/XRT, this indicates the range of obsIDs. The data for obsIDs 00031312035, 00031312036, and 00031312093 were excluded because they are not present in the Swift Master Catalog of Cen~A.\\
\footnotemark[$\dagger$]Start date of observations. For Swift/XRT, this is the start date of the first observation.\\
\footnotemark[$\ddagger$]End date of observations. For Swift/XRT, this is the end date of the last observation.\\
\footnotemark[$\S$]Exposure in units of ks. We adopt NuSTAR/FPMA, Suzaku/XIS 0, and XMM-Newton/EPIC-PIN exposures after data reduction. For Swift/XRT, this is the sum of the exposure of the observations in the duration.\\

\end{tabnote}
\end{table*}

\section{$\mathrm{Fe \ K\alpha}$ line reverberation mapping}
\label{rev_mapping}

To estimate the time lag of the reflection component, we compared the light curves of the direct and reflection components. The Swift/BAT ($15$--$50~\mathrm{keV}$) light curve served as the direct component, and the Fe~K$\alpha$ line fluxes as the reflection component. In subsection \ref{iron_line_flux_sub}, we carry out spectral analysis of multi-epoch observations from NuSTAR, Suzaku, XMM-Newton, and Swift/XRT to determine the Fe~K$\alpha$ line fluxes. In subsection \ref{transfer_sub}, we apply a transfer function method to determine the size of the reflector.

\subsection{Estimation of Fe~K$\alpha$ line fluxes}
\label{iron_line_flux_sub}

\begin{figure*}
 \begin{center}
  \includegraphics[width=12cm]{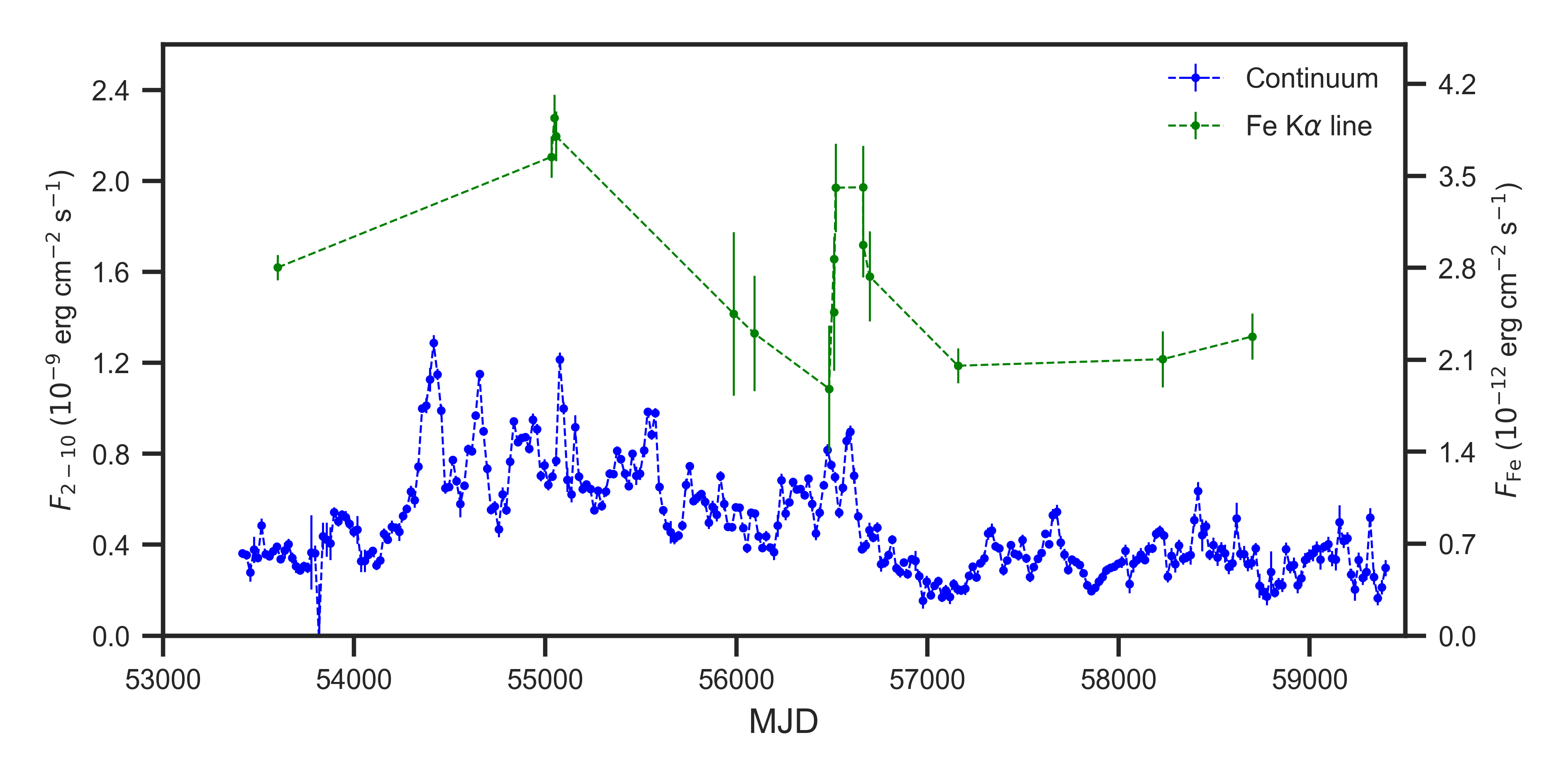}
 \end{center}
 \caption{Light curve of the hard X-ray continuum and that of the Fe~K$\mathrm{\alpha}$ line. The blue dots are unabsorbed continuum fluxes ($2$--$10~\mathrm{keV}$) obtained from Swift/BAT hard X-ray data ($15$--$50~\mathrm{keV}$), and the green dots are fluxes of the Fe~K$\alpha$ line inferred from the NuSTAR, Suzaku, XMM-Newton, and Swift/XRT data.}
 \label{lc_sw_fe}
\end{figure*}

To derive the Fe~K$\alpha$ line fluxes, we analyzed X-ray spectra from NuSTAR, Suzaku, XMM-Newton, and Swift/XRT. Spectra in the $4$--$10~\mathrm{keV}$ band ($4$--$9~\mathrm{keV}$ for Swift/XRT data) were modeled using a power-law and a Gaussian model: \\
\texttt{constant*phabs*(zphabs*cabs*zpowerlw + zgauss)}. The \texttt{constant} is the cross-normalization factor between the FI CCDs and the BI CCD, or between the FPMA and B. In the XMM-Newton and Swift/XRT spectral analyses, we fixed this constant at unity.
The \texttt{phabs} accounts for Galactic absorption, fixed at $2.35\times10^{20}~\mathrm{cm^{-2}}$ \citep{HI4PI2016}. The \texttt{zphabs} and \texttt{cabs} model the photon absorption and Compton scattering by the torus, respectively. Throughout this paper, we adopt a redshift for Cen~A of $z=0.0018$ \citep{Graham1978}. The \texttt{zpowerlw} represents the hard X-ray continuum emission from the nucleus of Cen~A, and \texttt{zgauss} models the Fe~K$\alpha$ fluorescence emission line, fixed at an energy of $6.4~\mathrm{keV}$ in the rest frame.
For the analysis of NuSTAR’s compromised energy resolution data and Swift/XRT's limited photon statistics data, the line width was set to 0. In the Swift/XRT spectral analysis, data were jointly fitted across the 2- or 3-month periods specified in table \ref{tab:datalist}, with all parameters, except for the normalization of \texttt{zpowerlw}, tied to consistent values across observations within these durations. Due to low photon counts in the Swift/XRT spectra, the cstat statistic was employed. For other datasets, the chi-squared statistic was used for fitting. The spectra and the estimated parameters are detailed in Appendix \ref{apx:spec_fit} (figure~\ref{fig:nustar_spec_gauss}--\ref{fig:swiftxrt_spec_gauss_2} and table~\ref{tb:fe_flux_fit_nustar}--\ref{tb:fe_flux_fit_swftxrt}). The light curve of the Swift/BAT and the obtained Fe~K$\alpha$ line fluxes are shown in figure~\ref{lc_sw_fe}.
To correct the flux of Fe~K$\alpha$ line flux measured by various instruments, we adjusted the cross-normalization factors based on \citet{Madsen2017}: multiplying by $4.0/(0.91+0.97+0.95+0.97)$ for Suzaku, $1.0/0.89$ for XMM-Newton, and $2.0/(1.01 + 1.08)$ for Swift/XRT data.

The Fe~K$\alpha$ line flux variations in Cen~A suggest the presence of a long-distance component, reflecting off distant material, and a short-distance component, reflecting from a sub-parsec-scale reflector. The ratio of the standard deviation to the mean flux of the Fe~K$\alpha$ line in Cen~A, $0.23$, was lower than that of the direct component, $0.45$. This discrepancy indicates the suppression of flux variation in the Fe~K$\alpha$ line due to a long-distance component. The reflection component emitted from materials along the line of sight exhibits no lag relative to the direct component. However, the reflection from the farther side of the reflector shows a lag of approximately $2r/c$, where $r$ is the distance from the source to the reflector and $c$ is the speed of light. The lag from other parts of the reflector varies between $0$ and $2r/c$. The overall reflection component comprises contributions from various lag components, ranging from $0$ to $2r/c$. Therefore, for larger values of $r$, the broader range of lag components leads to a suppression of flux variation in the light curve. The flux of the Fe~K$\alpha$ line decreased between the 2013 ($\mathrm{MJD}\simeq56510$) and 2015 ($\mathrm{MJD} \simeq 57160$) observations, paralleling a drop in the direct component flux ($\mathrm{MJD} \simeq 56650$; see figure \ref{lc_sw_fe}). This trend suggests the presence of a short-distance  ($\lesssim 500$~light-days) component.

\subsection{Transfer function method}
\label{transfer_sub}

To confirm the presence of both short- and long-distance components and to explore the geometry of the reflector emitting the Fe~K$\alpha$ line, we utilized the transfer function method. The essential procedure of this method is detailed in sub-subsection \ref{TF_method_subsub}. Sub-subsection \ref{single_top_hat_subsub} discusses the analysis assuming a top-hat transfer function, commonly employed in reverberation mapping studies (e.g., \cite{Pei2014}; \cite{Grier2017}; \cite{Noda2020}). Sub-subsection \ref{double_step_subsub} introduces a transfer function that integrates both short- and long-distance components.

\subsubsection{Procedure of transfer function method}
\label{TF_method_subsub}
In the transfer function method, the reflection component is modeled as the convolution of the direct component with the transfer function. This function quantifies how the light curve of the Fe~K$\alpha$ line responds if the input light curve of the direct component is a Dirac delta function. Here, we model the shape of the transfer function using a few parameters.

To calculate the convolution at the first Fe~K$\alpha$ data point, which occurred approximately $1.9\times10^2~{\mathrm{days}}$ after the first Swift/BAT data point, we estimated the light curve of the direct component prior to the beginning of Swift/BAT observations. We extended the Swift/BAT light curve to three times its original length using the following method. First, we created 2000 light curves, each of the same duration as the Swift/BAT light curve, by randomly changing the phase of its Fourier components. This ensured that the power spectra of these light curves matched that of the observed Swift/BAT light curve. We then paired these to generate 1000 extrapolation patterns, labeled from 0 to 999, which were utilized as the light curves preceding Swift/BAT observations. Despite potential flux discontinuities at the connection points, these are anticipated to have minimal effect on the main results of the analysis. This is because these flux gaps will be suppressed and smoothed out by the transfer function. The extrapolated light curves were linearly interpolated with a bin size of $5$~days for the numerical convolution calculations, treating the extrapolation label number as an additional free parameter.

We estimated the least $\chi^2$ value and best-fit parameters using the following procedure. The Fe~K$\alpha$ line fluxes were fitted with the convolution of the extrapolated Swift/BAT light curves and the transfer function by maximizing the evaluation function $ -\chi^2/2 =-\sum_{i}\left(f_i-y_i\right)^2/2\sigma_i^2 $, where $f_i$ represents the calculated convolution at data point $i$, and $y_i$ and $\sigma_i$ are the flux and its associated error at each data point, respectively. We employed the Markov Chain Monte Carlo (MCMC) method using \texttt{emcee} \citep{Foreman-Mackey2013}, with uniform prior distributions for the parameters as described in sub-subsections \ref{single_top_hat_subsub} and \ref{double_step_subsub}. The fitting was performed using \texttt{optimize.curve\_fit} in the Python Scipy package, initializing the parameters with values providing the least $\chi^2$ value in MCMC samples. Note that this analysis disregarded the error associated with the Swift/BAT light curve.

\subsubsection{Top-hat transfer function}
\label{single_top_hat_subsub}

We employed a top-hat transfer function:
\begin{equation}
\Psi(t) = \left\{
\begin{array}{cl}
\frac{s}{w} & (\tau - \frac{w}{2} \leq t < \tau + \frac{w}{2}), \\
0 & (\mathrm{otherwise}).
\end{array}
\right.
\label{one_top_hat_eq}
\end{equation}
The top-hat transfer function has three parameters: the lag of the reflection component ($\tau$), the width ($w$), and the area of the transfer function ($s$). This function is utilized in the JAVELIN algorithm (\cite{Zu2011}, \cite{Zu2013}) and is commonly applied in reverberation mapping studies (e.g., \cite{Pei2014}; \cite{Grier2017}; \cite{Noda2020}).

We conducted the MCMC with eight initial points, discarded the first $10^5$ steps, and ran $10^6$ steps for each chain. The prior distributions of the parameters were assumed that the uniform distribution within the ranges of $ \left[0.01~\mathrm{days},~5300~\mathrm{days}\right]$ for $\tau$, $\left[0~\mathrm{days},~2\tau\right]$ for $w$, and $\left[5\times 10^{-4},~5\times 10^{-2}\right]$ for $s$. The best-fit parameters obtained were $\tau = 4.9\times 10^3~\mathrm{days}$, $w = 8.8\times10^{2}~\mathrm{days}$, and $s = 5.7\times10^{-3}$, with a least $\chi^2$ value of $15.712$ for $12$ degrees of freedom.
Although a null hypothesis probability of $20\%$ is deemed acceptable, only a limited number of extrapolation patterns provided acceptable $\chi^2$ values: 4 out of 1000 patterns had a null hypothesis probability higher than $5.0\%$ in our MCMC samples. This occurred because the start time of the rise of the best-fit transfer function, $\tau - w/2 = 4.4\times 10^3$~days, was so large that the estimated Fe~K$\alpha$ line fluxes were mainly influenced by the extrapolated portions of the light curves for the direct component which were randomly generated. When the analysis was limited to MCMC samples with rise start times less than $3000~$days, the least $\chi^2$ value was $26$, indicating that the top-hat transfer function model was rejected at a $1 \times 10^{-2}$ significance level. These results indicate that the transfer function of this system is unlikely to be approximated by a top-hat function.

\subsubsection{Transfer function with short- and long-distance components}
\label{double_step_subsub}

To model the light curve of the Fe~K$\alpha$ line, we utilized a transfer function that includes both short- and long-distance components:
\begin{equation}
\Psi(t) = \left\{
\begin{array}{cl}
\frac{s\alpha}{2\tau_1} + \frac{s(1 - \alpha)}{2\tau_2} & (0\leq t < 2\tau_1),\\
\frac{s(1 - \alpha)}{2\tau_2} & (2\tau_1 \leq t < 2\tau_2), \\
0 & (\mathrm{otherwise}).
\end{array}
\right.
\label{two_top_hat_eq}
\end{equation}
This function has four parameters: the lag of the short-distance component ($\tau_1$), the lag of the long-distance component ($\tau_2$), the area of the transfer function ($s$), and the intensity ratio of the short-distance component to the total intensity ($\alpha$).
This transfer function simulates a reflector consisting of two spherical shells at different radii around the central source.
It serves as a simplified model for more complex geometrical shapes.

We conducted five independent MCMCs, each starting from ten initial points. We discarded the first $3 \times 10^5$ steps and continued for an additional $3\times10^5$ steps for each chain. The potential scale reduction factors\footnote{$R = \sqrt{V/W}$, where $V = \left(\left(n-1\right)/n\right)W + \left(1/n\right)B$, $W$ is the within-chain variance, $B$ is the between-chain variance, and $n$ is the length of each chain. $R$ is used to assess the convergence of MCMCs using the Gelman--Rubin diagnostic.} for each parameter were $1.025$, $1.003$, $1.005$, $1.002$, indicating well-converged MCMCs. Uniform priors were used within the ranges of $\left[0~\mathrm{days},~3000~\mathrm{days}\right]$ for $\tau_1$, $\left[\tau_1,~5300~\mathrm{days}\right]$ for $\tau_2$, $\left[1\times 10^{-3},~1\times 10^{-2}\right]$ for $s$, and $\left[0,~1\right]$ for $\alpha$. The analysis obtained an acceptable least $\chi^2$ value of $15.1$ for $11$ degrees of freedom. Unlike the top-hat function scenario, acceptable $\chi^2$ values were obtained across a broad range of extrapolated patterns: 403 out of 1000 patterns showed a null hypothesis probability greater than $5.0\%$ in the MCMC samples.

\begin{figure*}
 \begin{center}
  \includegraphics[width=140mm]{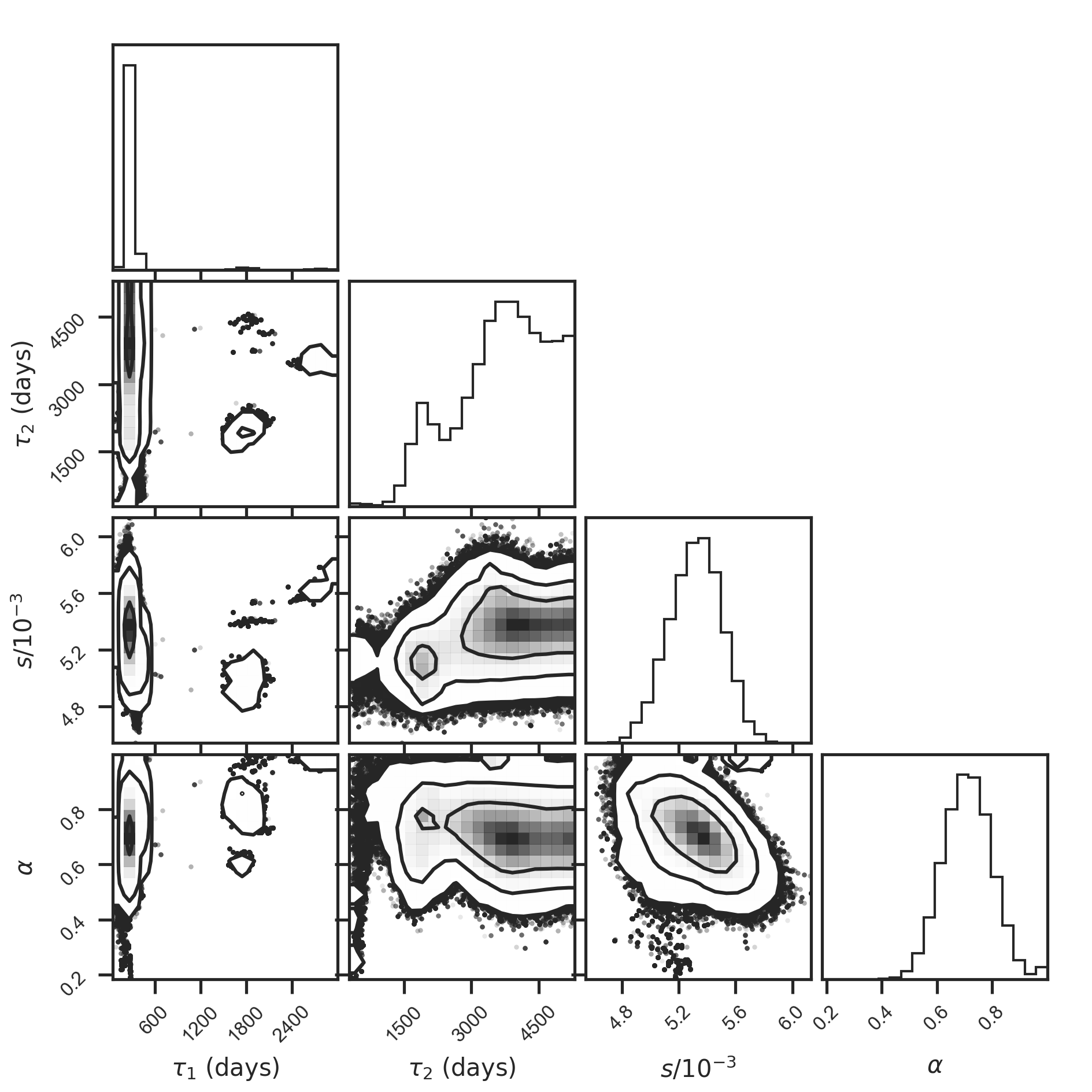}
 \end{center}
 \caption{Confidence contours among the transfer function described in equation (\ref{two_top_hat_eq}) parameters and the distributions of the parameters. The lines represent 68, 95, and 99.7\% confidence levels.}
 \label{contour_TF}
\end{figure*}

\begin{figure*}
 \begin{center}
  \includegraphics[width=140mm]{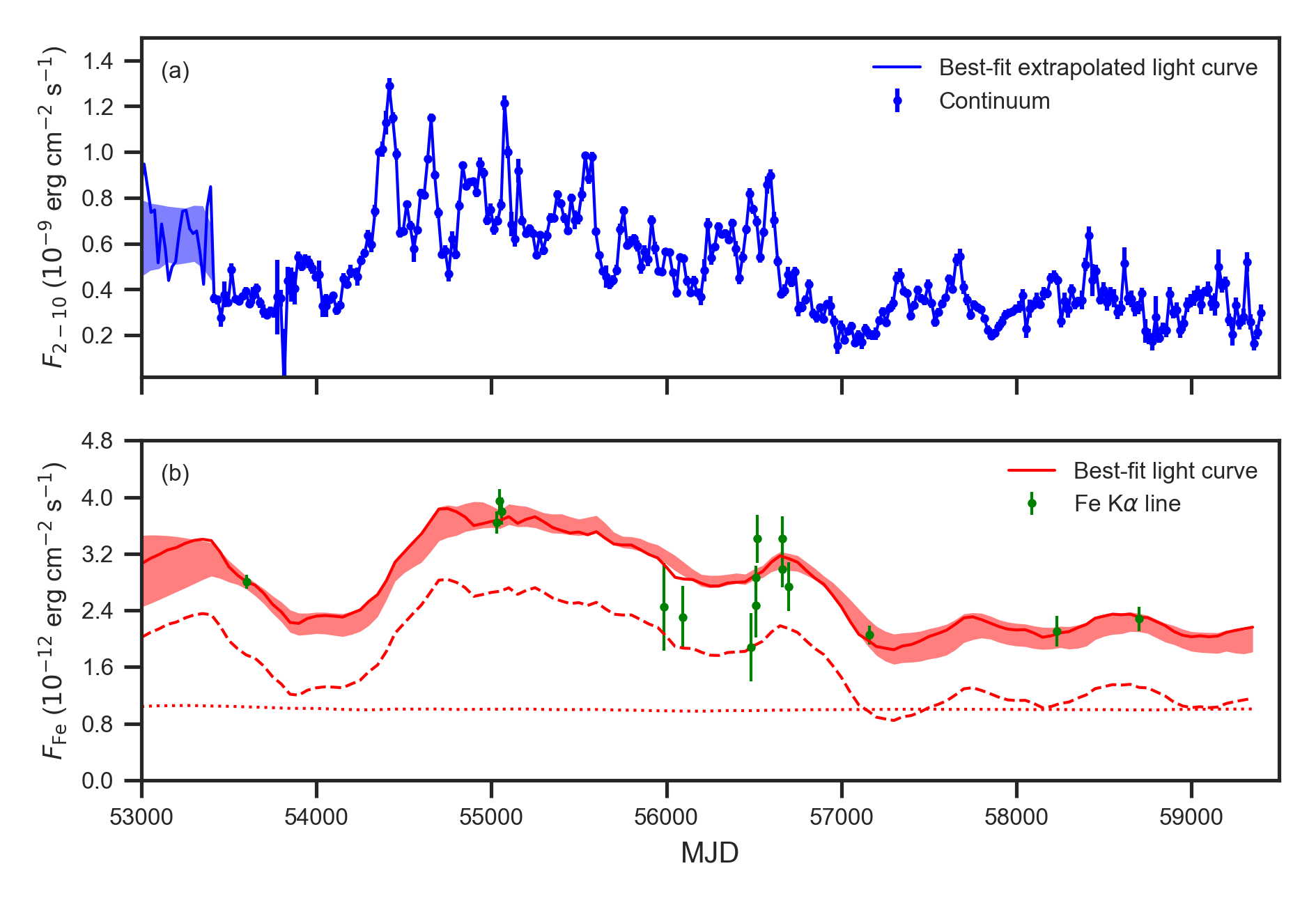}
 \end{center}
 \caption{(a) Light curve of the direct component obtained by extrapolating and interpolating the Swift/BAT light curve. The blue line is a part of the extrapolating and interpolating light curve calculated using the extrapolation pattern which provides the best-fit parameters. The shaded region represents $1\sigma$ intervals. The blue dots represent the light curve of Swift/BAT. (b) Light curve of the estimated Fe~K$\alpha$ line flux obtained from the convolution of the direct component and the best-fit transfer function. The red line represents the estimated Fe~K$\alpha$ line flux and the shaded region represents $1\sigma$ intervals. The red dashed and dotted lines represent short- and long-distance components, respectively. The green dots show the observed Fe~K$\alpha$ line fluxes.}
 \label{lc_sw_fe_fit}
\end{figure*}

The results are displayed in figure~\ref{contour_TF}, and the best-fit parameters and their errors are shown in table \ref{tf_para}. As shown in figure~\ref{lc_sw_fe_fit}, the transfer function with these parameters effectively captures the relation between the light curves of the direct component and the Fe~K$\alpha$ line. The lag of the short-distance component was $ 2.3_{- 0.3}^{+ 1.2} \times 10^2$ days ($0.19_{- 0.02}^{+ 0.10}~\mathrm{pc}$), and the lag of the long-distance component was greater than $2.1\times 10^3$ days ($>1.7~\mathrm{pc}$). The intensity ratio of the short-distance component was $0.56$--$0.85$, indicating that it contributes $56$\%--$85$\% of the total Fe~K$\mathrm{\alpha}$ line flux.

\begin{table*}
\tbl{Parameters for the transfer function.\footnotemark[$*$]}{%
\begin{tabular}{ccccc}
\hline
$\tau_1$\footnotemark[$\dagger$] & $\tau_2$\footnotemark[$\ddagger$] & $s$\footnotemark[$\S$] & $\alpha$\footnotemark[$\|$]&($\chi^2$, dof)\footnotemark[$\#$] \\
\hline
 $ 2.3_{- 0.3}^{+ 1.2} \times 10^{2} $ & $ > 2.1 \times 10^{3} $ &  $ 5.45_{- 0.43}^{+ 0.17}  \times 10^{-3} $ &  $ 0.63_{- 0.07}^{+ 0.22}  $  & (15.1, 11) \\
\hline
\end{tabular}}\label{tf_para}
  \begin{tabnote}
\footnotemark[$*$]The errors in the table are 90\% highest posterior density intervals estimated from the MCMC samples.\\
\footnotemark[$\dagger$]The first lag peak in days. \\
\footnotemark[$\ddagger$]The second lag peak in days. \\
\footnotemark[$\S$]The total area of the transfer function.\\
\footnotemark[$\|$]The ratio of the short-distance component area to the total area. \\
\footnotemark[$\#$] The extraction pattern labeled 617 gave the least $\chi^2$ value.\\
\end{tabnote}
\end{table*}

Figure \ref{lc_sw_fe_fit} shows that the light curve of the Fe~K$\alpha$ line can be decomposed into variable and constant components. To validate this hypothesis, we performed a similar analysis using MCMC, assuming that the Fe~K$\alpha$ line light curve is the sum of the convolution and a constant component:
\begin{equation}
F_{\mathrm{Fe}}(t) = \int F_{2-10}(t^{\prime})\tilde{\Psi}(t-t^{\prime})dt^{\prime} + C
\label{eq:fe_line_tophat_const}
\end{equation}
where $F_{\mathrm{Fe}}(t)$ and $F_{2-10}(t)$ are the light curves of the Fe~K$\alpha$ line and the direct component, respectively, and $C$ is a constant. The transfer function $\tilde{\Psi}(t)$ is given by
\begin{equation}
\tilde{\Psi}(t) = \left\{
\begin{array}{cl}
\frac{\tilde{s}}{2\tau_1} & (0\leq t < 2\tau_1),\\
0 & (\mathrm{otherwise}).
\end{array}
\right.
\label{eq:tranfer_constant}
\end{equation}
Equation (\ref{eq:fe_line_tophat_const}) corresponds to the case where the contribution from the long-distance component in equation (\ref{two_top_hat_eq}) to the Fe~K$\alpha$ line flux, $\frac{s(1 - \alpha)}{2\tau_2} \int_{t-2\tau_2}^{t}dt^{\prime}F_{2-10}(t^{\prime})$, can be regarded as a constant.

We conducted MCMCs similar to the analysis using the transfer function in equation (\ref{two_top_hat_eq}). This time, each MCMC started from eight initial points. The analysis obtained an acceptable least $\chi^2$ value of $15.5$ for $12$ degrees of freedom, and 437 out of 1000 extrapolation patterns showed a null hypothesis probability greater than $5.0\%$ in the MCMC samples. The obtained value of the lag was $2.4_{- 0.4}^{+ 1.0} \times 10^2$ days, which is consistent with the inferred value from the analysis using equation (\ref{two_top_hat_eq}). The inferred value of $\tilde{s}$ was $3.55_{- 0.47}^{+ 0.74} \times 10^{-3}$ and $C$ was $9.5_{- 4.6}^{+ 2.7} \times 10^{-13}~\mathrm{erg~cm^{-2}~s^{-1}}$.

\section{Spectral analysis}
\label{spectral}

To investigate the Compton thickness of the reflector and define the properties of the primary continuum component, we analyzed spectral data from four epochs of NuSTAR and six epochs of Suzaku. The transfer function derived from reverberation mapping was utilized in this analysis. As NuSTAR, along with Suzaku/XIS and HXD-PIN, covers both the energy range of the Fe~K$\alpha$ line and the reflection continuum, these instruments facilitate the determination of reflector properties. We used XSPEC version 12.14.0 integrated into HEASoft v6.33. We fitted the NuSTAR data spanning $4$ to $78~\mathrm{keV}$ and the Suzaku data from $4$ to $10~\mathrm{keV}$ (XIS) and $15$ to $50~\mathrm{keV}$ (HXD-PIN) using a model that incorporates both the direct and reflection components. The normalization of the reflection component was calculated using the transfer function estimated in sub-subsection \ref{double_step_subsub} to account for the lag of the reflection component. The XClumpy model \citep{Tanimoto2019} was selected as the reflection model because it accommodates both Compton-thin and Compton-thick scenarios. Variability in the absorbing column density of Cen~A supported the application of the clumpy torus model (\cite{Rothschild2011}; \cite{Rivers2011}).
While the XClumpy model's transfer function might not perfectly match with the transfer function described in equation (\ref{two_top_hat_eq}), adjusting parameters such as the inner and outer radii of the torus, inclination angle, and clump radial distribution to better match the shape of the transfer function described in equation (\ref{two_top_hat_eq}) could be possible. However, achieving a consistent model that fits both the light curves and spectra is beyond the scope of this paper. As reported in \citet{Furst2016}, the optically thick disk reflection model, \texttt{pexmon}, was found to be inadequate for modeling the NuSTAR spectrum, where the Fe~K$\alpha$ line is observed but the Compton hump is not clearly seen. 

We adopted the following model: \\
\texttt{constant*phabs*(zphabs*cabs*zcutoffpl \\
+ atable\{xclumpy\_v01\_RC.fits\} \\
+ atable\{xclumpy\_v01\_RL.fits\})}. \\
The initial factor adjusts the cross-normalization factors between FPMA and FPMB (NuSTAR), between XIS-FI and XIS-BI (Suzaku), or between XIS-FI and HXD-PIN (Suzaku). The Galactic hydrogen column density in the second factor was set at $2.35\times 10^{20}~\mathrm{cm^{-2}}$ \citep{HI4PI2016}. In accordance with the XClumpy model \citep{Tanimoto2020}, the cutoff energy for the direct component, \texttt{zcutoffpl}, was fixed at $370~\mathrm{keV}$, reflecting the typical value for low-Eddington AGNs in BAT samples \citep{Ricci2018}. The hydrogen column density along the line of sight was allowed to vary.

We used a clumpy torus model (XClumpy: \cite{Tanimoto2019}), in which the clump distribution follows a power-law in the radial direction and a Gaussian distribution in the elevation direction. XClumpy estimates both reflection continuum (\texttt{atable\{xclumpy\_v01\_RC.fits\}}) and line emissions (\texttt{atable\{xclumpy\_v01\_RL.fits\}}) from a given direct component spectrum. This model comprises six parameters: the cutoff energy, normalization, and photon index of the input spectrum; inclination angle; the hydrogen column density along the equatorial plane ($N_{\mathrm{H}}^{\mathrm{Equ}}$); and the torus angular width ($\sigma$). We set the inclination angle to 30 degrees, in line with the constraint on the angle between the VLBI jet and the line of sight \citep{Muller2014}. The parameters related to torus geometry, $N_{\mathrm{H}}^{\mathrm{Equ}}$ and $\sigma$, were tied across the all spectra. The light curve analysis in section \ref{rev_mapping} revealed a lag in the reflection component relative to the direct component, prompting adjustments in our spectral analysis. To take this time lag into account, we calculated the input spectrum of XClumpy as follows. The photon index was fixed at $1.8$, a typical value for the direct component in Cen A, with the cutoff energy set at $370~\mathrm{keV}$. We determined the normalization of the input direct component, $\mathrm{photons~cm^{-2}~s^{-1}~keV^{-1}}$ at $1~\mathrm{keV}$, $\mathrm{norm}_\mathrm{input}(t)$, using the following steps: First, we estimated the unabsorbed fluxes of the input direct component ($2$--$10~\mathrm{keV}$), $F_{2-10}(t)$, by using $F_{2-10}(t)=\int d\tau \hat{\Psi}(\tau)C(t-\tau)$, where $C(t)$ is the extrapolated Swift/BAT light curve, and
\begin{equation}
\hat{\Psi}(t) = \left\{
\begin{array}{cl}
\frac{\hat{\alpha}}{2\hat{\tau_{1}}} + \frac{1 - \hat{\alpha}}{2\hat{\tau_{2}}} & (0\leq t < 2\hat{\tau_{1}}),\\
\frac{1 - \hat{\alpha}}{2\hat{\tau_{2}}} & (2\hat{\tau_{1}} \leq t < 2\hat{\tau_{2}}), \\
0 & (\mathrm{otherwise}),
\end{array}
\right.
\label{two_top_hat_best_eq}
\end{equation}
is the estimated transfer function, with $\hat{\tau_{1}}$, $\hat{\tau_{2}}$, and $\hat{\alpha}$ representing the best-fit parameters from table \ref{tf_para}.
We set $s$ to unity to estimate the flux of the input direct component spectrum, $F_{2-10}(t)$, instead of the Fe~K$\mathrm{\alpha}$ line. The $F_{2-10}(t)$ was then converted into $\mathrm{norm}_\mathrm{input}(t)$, assuming a photon index of $1.8$ and a cutoff energy of $370~\mathrm{keV}$.

\begin{figure*}
 \begin{center}
  \includegraphics[width=8cm]{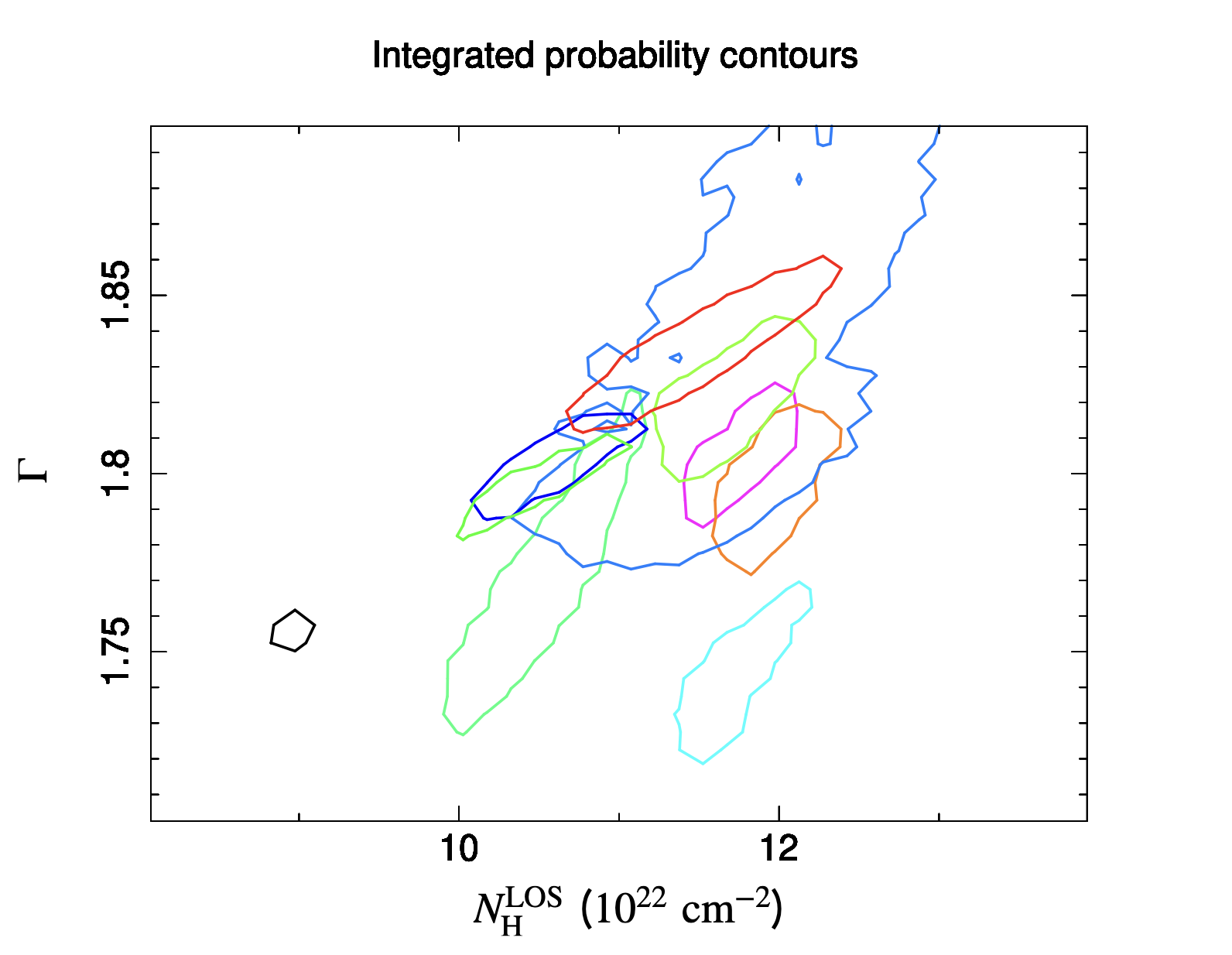}
 \end{center}
 \caption{Integrated probability contours of 90\% confidence level between the hydrogen column density along the line of sight in units of  $\rm
10^{22}\ \mathrm{cm}^{-2}$ and the photon index fitted model with XClumpy model. The contours estimated using $10^5$ MCMC samples: a chain with $10$ walkers and a total length of $10^5$.}
 \label{contour_nH_gamma}
\end{figure*}

 All spectra were well explained by this model, showing $\chi^2/\mathrm{dof} = 1.020$ for $18328$ degrees of freedom. Table \ref{xclumpy_paras} lists the best-fit parameters with the XClumpy model, and figures~\ref{fig:spec_nustar1}--\ref{fig:spec_suzaku2} in Appendix \ref{apx:spec_fit} display the folded X-ray spectra and best-fit model components. The value of $N_{\mathrm{H}}^{\mathrm{Equ}} = 3.14_{-0.74}^{+0.44} \times 10^{23}~ \mathrm{cm}^{-2}$ was less than $10^{24}~\mathrm{cm^{-2}}$, indicating that the torus was Compton-thin. Figure~\ref{contour_nH_gamma} shows the integrated probability contour between the hydrogen column density along the line of sight and the photon index, highlighting the degeneracy between these parameters. The change in the photon index ($\Delta \Gamma$) during the period from 2005 to 2019 was less than approximately 0.1, although the flux in 2013 was approximately four times larger than in 2015.

\begin{table*}
\tbl{best-fit parameters with XClumpy model\footnotemark[$*$].}{%
\begin{tabular}{lcccccccc}
\hline
ObsID&$N_{\mathrm{H}}^{\mathrm{LOS}}$\footnotemark[$\dagger$] & $\Gamma$\footnotemark[$\ddagger$] & norm\footnotemark[$\S$] & C$_{1}$\footnotemark[$\|$] & C$_{2}$\footnotemark[$\#$]  & $N_{\mathrm{H}}^{\mathrm{Equ}}$\footnotemark[$**$] & $\sigma$\footnotemark[$\dagger$] & ($\chi^2$, dof)\\
\hline
60001081002 & $ 8.92 \pm 0.19 $ & $ 1.755 \pm 0.005 $ & $ 0.221 \pm 0.003 $ & $ 1.030 \pm 0.003 $ & & & &  \\
60101063002 & $ 11.56_{-0.67}^{+0.70} $  & $ 1.840_{-0.018}^{+0.019} $  & $ 0.060_{-0.003}^{+0.004} $ & $ 1.019 \pm 0.008 $ & & & &  \\
60466005002 & $ 10.57_{-0.47}^{+0.48} $ & $ 1.797 \pm 0.013 $ & $ 0.120 \pm 0.005 $ & $ 1.009 \pm 0.006 $ & & & &  \\
10502008002 & $ 10.52_{-0.46}^{+0.48} $ & $ 1.801 \pm 0.013 $ & $ 0.105 \pm 0.004 $ & $ 1.017 \pm 0.006 $ & & & &  \\
100005010 & $ 11.64_{-0.35}^{+0.34} $  & $ 1.735 \pm 0.021 $  & $ 0.109 \pm 0.005 $  & $ 0.984 \pm 0.006 $  & $ 0.975 \pm 0.020 $ & & &  \\
704018010 & $ 11.59_{-0.32}^{+0.34} $  & $ 1.800 \pm 0.019 $  & $ 0.190_{-0.008}^{+0.009} $  & $ 0.989 \pm 0.005 $  & $ 1.332 \pm 0.025 $ & & &  \\
704018020 & $ 11.82_{-0.38}^{+0.40} $  & $ 1.795 \pm 0.023 $  & $ 0.175_{-0.009}^{+0.010} $  & $ 0.976 \pm 0.006 $  & $ 1.251_{-0.027}^{+0.028} $ & & &  \\
704018030 & $ 11.63_{-0.38}^{+0.39} $  & $ 1.820 \pm 0.022 $  & $ 0.191 \pm 0.010 $  & $ 0.973 \pm 0.006 $  & $ 1.360 \pm 0.029 $ & & &  \\
708036010 & $ 10.64_{-0.61}^{+0.62} $  & $ 1.778 \pm 0.039 $  & $ 0.213_{-0.018}^{+0.020} $  & $ 0.896 \pm 0.009 $  & $ 1.148_{-0.043}^{+0.045} $ & & &  \\
708036020 & $ 11.85_{-1.11}^{+1.13} $  & $ 1.839_{-0.070}^{+0.071} $  & $ 0.124_{-0.019}^{+0.022} $  & $ 0.888_{-0.015}^{+0.016} $  & $ 1.167_{-0.075}^{+0.080} $ & & &  \\
All & & & & & & $ 31.4_{-7.4}^{+4.4} $ & $ 19.1_{-1.5}^{+8.5} $ &(18691, 18328) \\
\hline
\end{tabular}}\label{xclumpy_paras}
  \begin{tabnote}
\footnotemark[$*$] The uncertainties in the table represent the 90\% confidence intervals.\\ 
\footnotemark[$\dagger$]Hydrogen column density along the line of sight in units of  $\rm
10^{22}\ \mathrm{cm}^{-2}$. \\
\footnotemark[$\ddagger$]The photon index of the direct component. \\
\footnotemark[$\S$]The normalization at 1 keV in units of $\mathrm{photons~cm^{-2}~s^{-1}~keV^{-1}}$.\\ \footnotemark[$\|$]Cross-normalization factors between FPMA and FPMB (NuSTAR) or XIS-FI and XIS-BI (Suzaku). \\
\footnotemark[$\#$]Cross-normalization factors between XIS-FI and HXD-PIN (Suzaku). \\
\footnotemark[$**$]Hydrogen column density along the equatorial plane in units of $\rm
10^{22}\ \mathrm{cm}^{-2}$. \\
\footnotemark[$\dagger\dagger$]Torus angular width in units of degrees. 
\end{tabnote}
\end{table*}

\section{Discussion}
\label{discussion}
Through X-ray reverberation mapping using data from Swift/BAT, NuSTAR, Suzaku, XMM-Newton, and Swift/XRT, we found that the time lag between the direct and reflection components displayed two distinct scales: $2.3_{- 0.3}^{+ 1.2} \times10^2~\mathrm{days}$ and $>2.1\times 10^3~\mathrm{days}$. The multi-epoch spectral analysis demonstrated that the hard X-ray spectra of Cen~A could be explained by the Compton-thin reflection model, accounting for the reflection component’s time lag.

We revealed that the transfer function for the Fe~K$\alpha$ line could be approximated by the transfer function in equation (\ref{two_top_hat_eq}), with lags of $2.3_{- 0.3}^{+ 1.2} \times 10^2~\mathrm{days}$ and $>2.1\times 10^3~\mathrm{days}$. Multiplying by the speed of light, these correspond to distances of $0.19_{- 0.02}^{+ 0.10}~\mathrm{pc}$ ($\sim 10^5 R_{\mathrm{g}}$) and $>1.7~\mathrm{pc}$ ($\sim 10^6R_{\mathrm{g}}$), where $R_{\mathrm{g}} = GM_{\mathrm{BH}}/c^2$. The time scale for the short-distance component aligns with the width of the Fe~K$\alpha$ line (e.g., \cite{Evans2004}; \cite{Markowitz2007}; \cite{Shu2011}). \citet{Evans2004} analyzed Chandra/HETG data, constraining the FWHM velocity ($v_{\mathrm{FWHM}}$) to between $1000~\mathrm{km~s^{-1}}$ and $3000~\mathrm{km~s^{-1}}$.
If we assume isotropic velocity distribution ($\left<v^2\right> = 3 \left< v^2_{\mathrm{LOS}}\right>$, where $\left<v^2_{\mathrm{LOS}}\right>$ is the line of sight velocity dispersion and $\left<v^2\right>$ is the velocity dispersion),  $v_{\mathrm{FWHM}}^2 = 4 \left< v^2_{\mathrm{LOS}}\right>$, and the Keplerian motion ($GM_{\mathrm{BH}} = r\left<v^2\right>$, with $r$ representing the distance from the black hole to the emitting gas of the Fe~K$\alpha$ line), then the relation between $r$ and $v_{\mathrm{FWHM}}$ is expressed as $r = 4GM_{\mathrm{BH}}/3v_{\mathrm{FWHM}}^{2}$ \citep{Netzer1990}. This leads to distances of $0.2~\mathrm{pc}$ and $0.03~\mathrm{pc}$ for $v_{\mathrm{FWHM}} = 1000~\mathrm{km~s^{-1}}$ and $3000~\mathrm{km~s^{-1}}$, respectively.
\citet{Shu2011} provided similar constraints for $v_{\mathrm{FWHM}}$. \citet{Markowitz2007}, analyzing Suzaku spectra, suggested $v_{\mathrm{FWHM}}< 2500~\mathrm{km~s^{-1}}$, consistent with our short-distance component findings. 

The scale of the short-distance component matches the size of Cen~A's torus, estimated at a sub-parsec scale. Using the bolometric luminosity value of $10^{43}~\mathrm{erg~s^{-1}}$ \citep{Whysong2004} with formula (1) from \citet{Nenkova2008b}, the dust sublimation radius is estimated at approximately $ 0.04~\mathrm{pc}$, potentially representing the inner radius of the torus. Infrared data analysis with a clumpy torus model suggested an inner radius of $0.021^{+0.002}_{-0.002}~\mathrm{pc}$ and an outer radius of $0.4^{+0.1}_{-0.1}~\mathrm{pc}$ \citep{Ichikawa2015}. \citet{Rivers2011} estimated the minimum torus size as approximately $0.1~\mathrm{pc}$ from the maximum column density and duration of an occultation event. The short-distance component scale from our analysis aligns with these estimates. 

\citet{Andonie2022} studied the origin of Fe~K$\alpha$ lines in bright nearby AGNs, including Cen~A, by comparing the light curves of the continuum and the Fe~K$\alpha$ line from spectral analyses of Chandra data. Their constraint for Cen~A is $<0.039~\mathrm{pc}$, about an order of magnitude smaller than our results. However, their results might stem from artifacts. The light curves obtained through their spectral analysis differ from those from Swift/BAT. For instance, their continuum light curve varied by a factor of $\sim 5$--$10$ within a short time scale, $\sim 1$ month. This behavior could be partially caused by the photon pile-up effect in Chandra data. They extracted spectra from an annulus region with an inner radius of $3$~arcsec and an outer radius of $5$~arcsec, which might not sufficiently avoid pile-up. Our analysis of a Chandra observation (obsID 7800) revealed that the flux increased when a $5^{\prime\prime}$--$7^{\prime\prime}$ annulus was used for spectral extraction.

In contrast, \citet{Furst2016} noted that the stable Fe~K$\alpha$ line flux (e.g., \cite{Rothschild2006}; \cite{Rothschild2011}) indicated that an emitter located $10$~lt-yr (approximately $3~\mathrm{pc}$) or more from the core, which aligns with the time scale of our long-distance component. Given that $\alpha = 0.63_{- 0.07}^{+ 0.22}$ and $\tau_1 = 2.3_{- 0.3}^{+ 1.2} \times 10^2$~days, precise Fe~K$\alpha$ line flux measurements with errors below $\sim 10$\% and significant hard X-ray variability over a $\sim 200$-day time scale are required to detect the short-distance component.

The transfer function in equation (\ref{two_top_hat_eq}), with its inferred parameters, suggests that the reflection component’s emission regions are distinctly located at approximately $ 0.19~\mathrm{pc}$ (dust torus scale) and more than $>1.7~\mathrm{pc}$ (e.g., circumnuclear disk scale $\sim 10^2~\mathrm{pc}$; \cite{Espada2009}) from the SMBH. However, it is plausible that the emission region extends continuously from approximately $0.19~\mathrm{pc}$ to a parsec scale. Our ability to determine the transfer function's shape is restricted due to the limited number of Fe~K$\alpha$ line flux data points, necessitating a simplified transfer function with few parameters. Additionally, while the transfer function in equation (\ref{two_top_hat_eq}) offers one explanation, it might not be the only model that aligns with the data. Therefore, we cannot definitively conclude whether the Fe~K$\alpha$ line originates from two separated reflectors with different scales or from a reflector extending continuously over the inferred scales.

Our analysis faces several limitations. Primarily, the typical intervals between the Fe~K$\mathrm{\alpha}$ line flux data, ranging from several hundred to a thousand days, prevent us from imposing constraints on very short-distance components, such as those spanning only $10$~days. Therefore, the lower limit of the short-distance component is not reliably established. Microcalorimeter missions, such as the X-Ray Imaging and Spectroscopy Mission (XRISM; \cite{Tashiro2022}), will allow us to better constrain the emission regions from the Fe~K$\alpha$ line profile. In addition, Swift/BAT data are only available from February 2005, limiting our ability to examine very long components with time scales exceeding several thousand days.

Previous works have shown that Compton-thick reflection models can explain the hard X-ray spectra of Cen~A (e.g., \cite{Fukazawa2011b}; \cite{Burke2014}). However, analyses of the NuSTAR spectrum in 2013 suggested that the reflector is Compton-thin (\cite{Furst2016}; \cite{Ogawa2021}). These previous studies did not consider the time lag of the reflection component in spectral analysis, and \citet{Fukazawa2011b} did not account for the photon pile-up effect in their Suzaku/XIS data. Our spectral analysis, which accounted for these effects, revealed that the hard X-ray spectra could be modeled with a power-law component and the reflection component from the reflector with hydrogen column density along the equatorial plane was less than $10^{24}~\mathrm{cm^{-2}}$. This finding supports the notion that Compton-thin material originates the Fe~K$\alpha$ line.

\section{Conclusions}
\label{conclusion}

We analyzed the light curves of the direct and reflection components of Cen~A using archival data from NuSTAR, Suzaku, XMM-Newton, and Swift. We found that a top-hat transfer function is unlikely, although it is commonly employed in reverberation mapping studies. Instead, a transfer function featuring short- and long-distance components adequately explains these light curves, with inferred distances of $0.19_{- 0.02}^{+ 0.10}~\mathrm{pc}$ and $> 1.7~\mathrm{pc}$. The short-distance component contributes $56$\%--$85$\% of the total Fe~K$\alpha$ line flux.

In addition, we examined spectral data from four NuSTAR epochs and six Suzaku epochs, considering the lag of the reflection component. The analysis revealed that the core of Cen~A is surrounded by Compton-thin material, with an equatorial hydrogen column density of  $N_{\mathrm{H}}^{\mathrm{Equ}} = 3.14_{-0.74}^{+0.44} \times 10^{23}~ \mathrm{cm}^{-2}$. These results suggest that the Fe~K$\alpha$ emission may originate from either a Compton-thin dust torus located at sub-parsec scale and materials farther away, or a Compton-thin torus that extends continuously from sub-parsec to parsec scales.

\begin{ack}
We thank the anonymous reviewer for their helpful comments. This work was supported financially by Forefront Physics and Mathematics Program to Drive Transformation (FoPM), a World-leading Innovative Graduate Study (WINGS) Program, the University of Tokyo (T.I.), and by Research Fellowships of the Japan Society for the Promotion of Science (JSPS) for Young Scientists (T.I.). Funding was also provided by JSPS Grants-in-Aid for Scientific Research (KAKENHI) Grant Numbers JP23KJ0780 (T.I.), JP22K18277, JP22H00128 (H.O.), and JP23H01211 (A.B.). A.T. is supported by the Kagoshima University postdoctoral research program (KU-DREAM). H.O. is supported by Toray Science and Technology Grant 20-6104. Y.I. is supported by JSPS KAKENHI Grant Numbers JP18H05458, JP19K14772, and JP22K18277. Additional support came from the World Premier International Research Center Initiative (WPI), MEXT, Japan.
\end{ack}

\appendix
\section{Results of spectral analysis}
\label{apx:spec_fit}

\begin{figure*}
 \begin{center}
  \includegraphics[width=12cm]{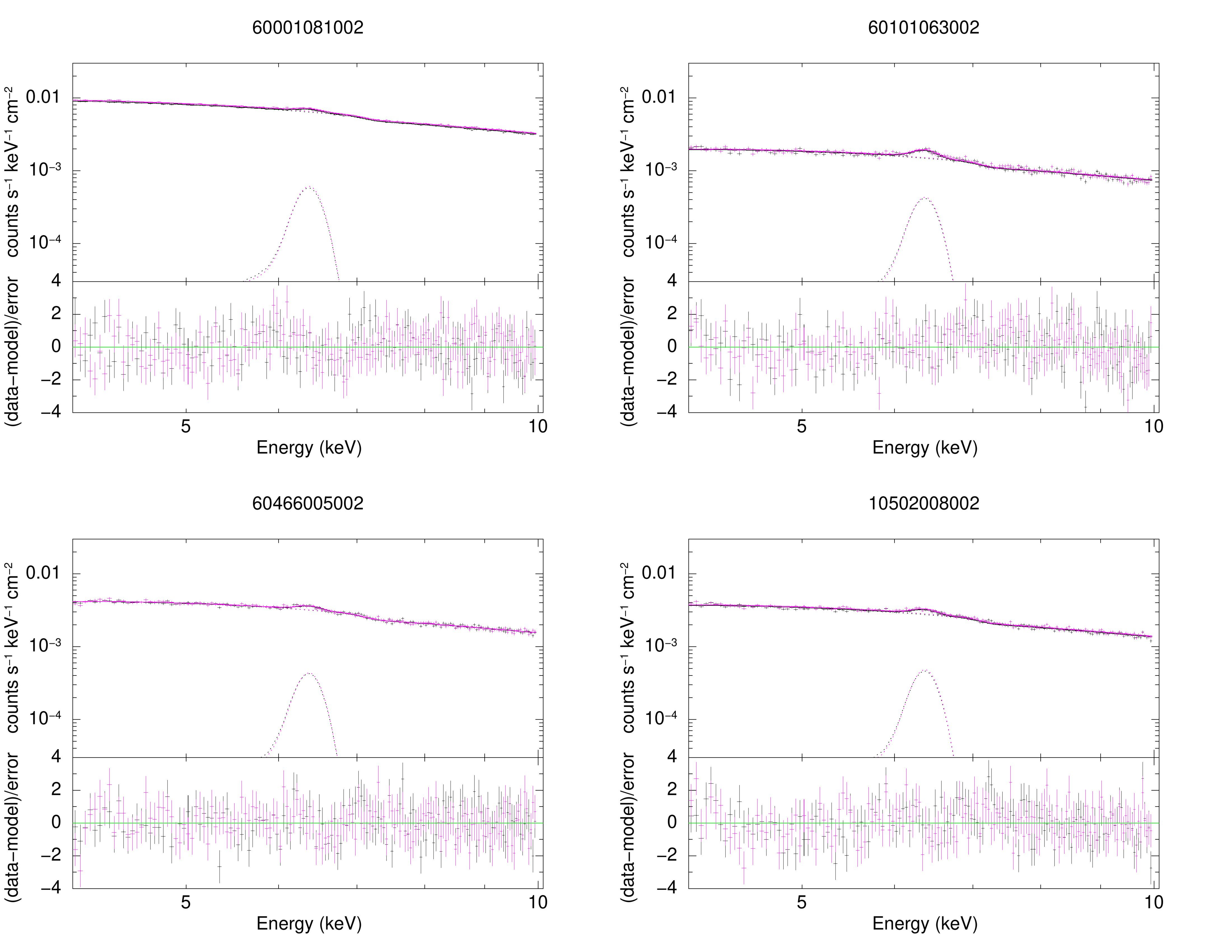}
 \end{center}
 \caption{Folded X-ray spectra fitted with the model \texttt{constant*phabs*(zphabs*cabs*zpowerlw + zgauss)}. The black and magenta crosses represent data from NuSTAR/FPMA and FPMB, respectively. The solid lines show the best-fit model, while the black and magenta dotted lines indicate its components. The lower panels display residuals.}
 \label{fig:nustar_spec_gauss}
\end{figure*}

\begin{figure*}
 \begin{center}
  \includegraphics[width=12cm]{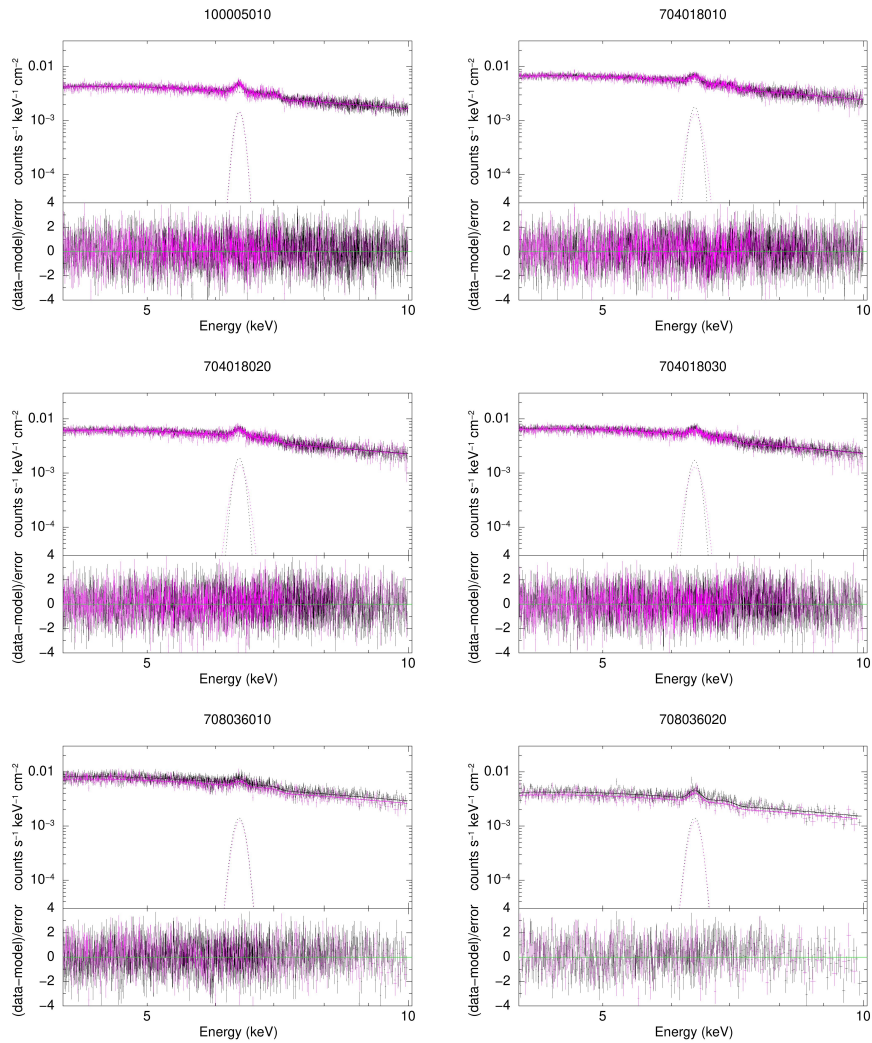}
 \end{center}
 \caption{Folded X-ray spectra fitted with the model \texttt{constant*phabs*(zphabs*cabs*zpowerlw + zgauss)}. The black and magenta crosses represent data from Suzaku/XIS-FI and XIS-BI, respectively. The solid lines show the best-fit model, while the black and magenta dotted lines indicate its components. The lower panels display residuals.}
 \label{fig:suzaku_spec_gauss}
\end{figure*}

\begin{figure*}
 \begin{center}
  \includegraphics[width=12cm]{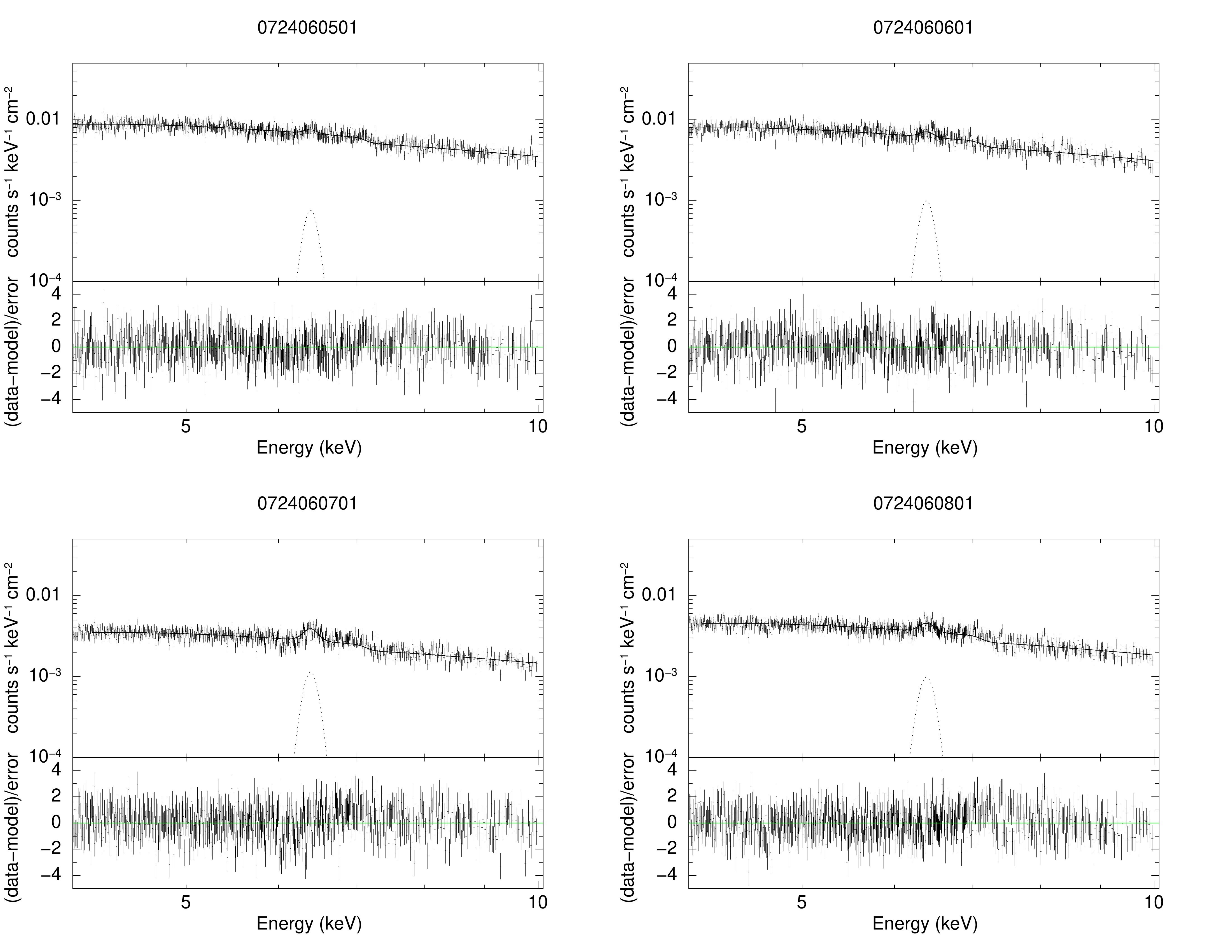}
 \end{center}
 \caption{Folded X-ray spectra fitted with the model \texttt{constant*phabs*(zphabs*cabs*zpowerlw + zgauss)}. The black crosses represent data from XMM-Newton/EPIC-PN. The solid lines show the best-fit model, while the black dotted lines indicate its components. The lower panels display residuals.}
 \label{fig:xmm_spec_gauss}
\end{figure*}

\begin{figure*}
 \begin{center}
  \includegraphics[width=12cm]{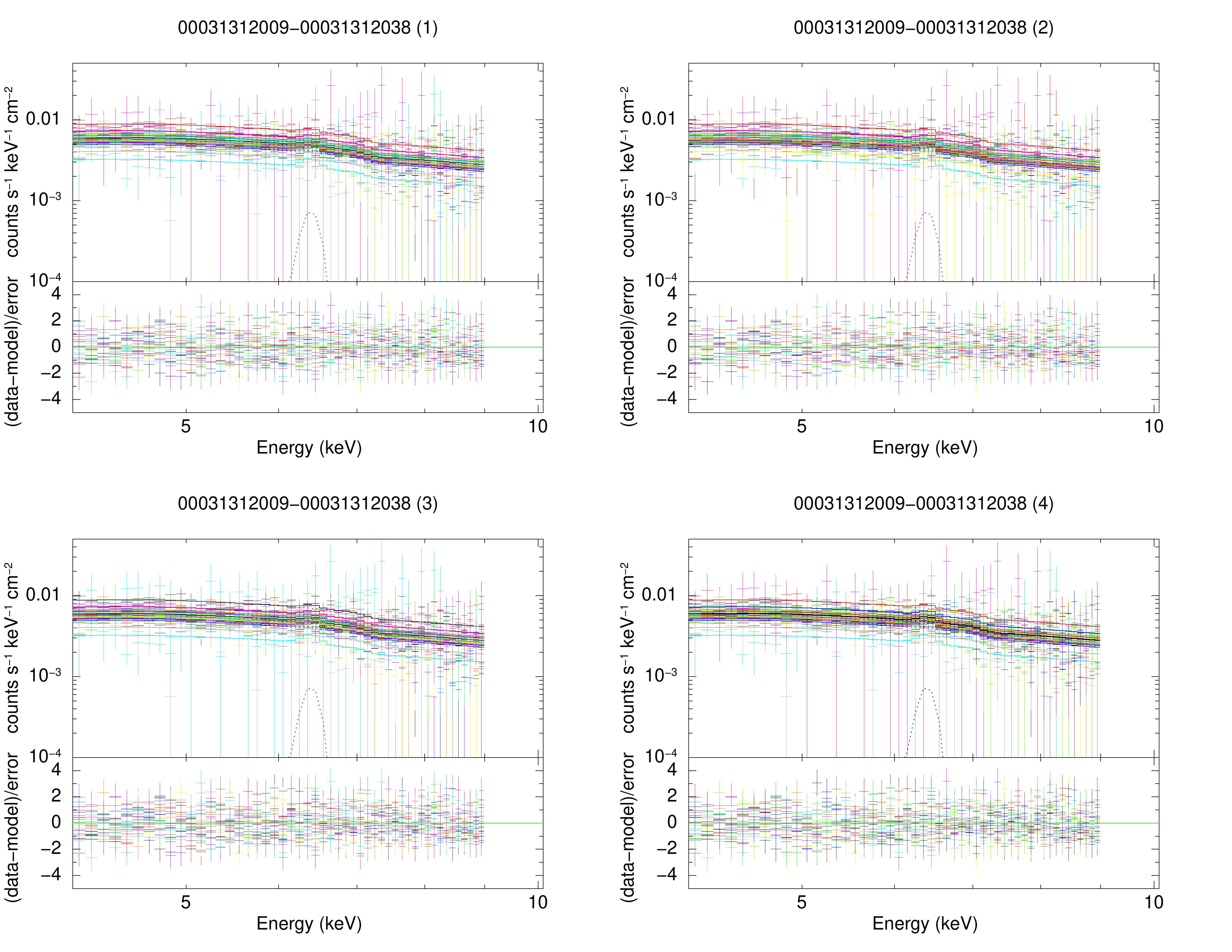}
 \end{center}
 \caption{Folded X-ray spectra fitted with the model \texttt{constant*phabs*(zphabs*cabs*zpowerlw + zgauss)}. The crosses represent data from Swift/XRT. The solid lines show the best-fit model, while the dotted lines indicate its components. The lower panels display residuals. For visual clarity, the data were rebinned and plotted in four separate panels.}
 \label{fig:swiftxrt_spec_gauss_1}
\end{figure*}

\begin{figure*}
 \begin{center}
  \includegraphics[width=12cm]{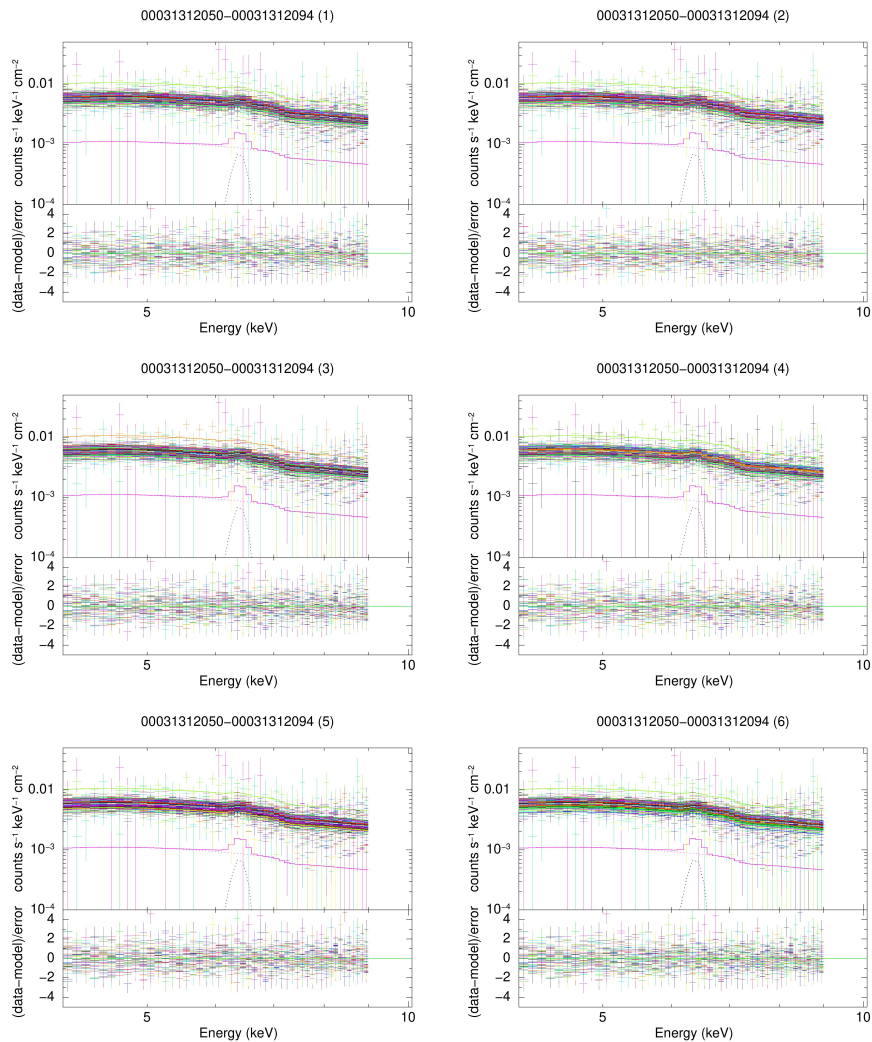}
 \end{center}
 \caption{Folded X-ray spectra fitted with the model \texttt{constant*phabs*(zphabs*cabs*zpowerlw + zgauss)}. The crosses represent data from Swift/XRT. The solid lines show the best-fit model, while the dotted lines indicate its components. The lower panels display residuals. For visual clarity, the data were rebinned and plotted in six separate panels.}
 \label{fig:swiftxrt_spec_gauss_2}
\end{figure*}

The results of the spectral analysis in subsection \ref{iron_line_flux_sub}, including the spectra and estimated parameters, are presented in figure~\ref{fig:nustar_spec_gauss} to \ref{fig:swiftxrt_spec_gauss_2} and table~\ref{tb:fe_flux_fit_nustar} to \ref{tb:fe_flux_fit_swftxrt}. 
The X-ray spectra fitted with the XClumpy model in section \ref{spectral} are displayed in figure \ref{fig:spec_nustar1} to \ref{fig:spec_suzaku2}, along with the best-fit model components.

\begin{table*}
\tbl{Best-fit parameters for NuSTAR spectral analysis with power-law and Gaussian model\footnotemark[$*$].}{%
\begin{tabular}{lccccccc}
\hline
obsID &$N_{\mathrm{H}}^{\mathrm{LOS}}$\footnotemark[$\dagger$]& $\Gamma$\footnotemark[$\ddagger$]& norm\footnotemark[$\S$] & $\log_{10}F_\mathrm{Fe}$\footnotemark[$\|$] & $\sigma_\mathrm{Fe}$\footnotemark[$\#$] & C$_{\mathrm{FPMB}}$\footnotemark[$**$] &($\chi^2$, dof)\\
\hline
60001081002 & $ 8.87 \pm 0.18 $  & $ 1.763 \pm 0.009 $  & $ 0.228 \pm 0.005 $  & $ -11.543_{-0.025}^{+0.024} $  & $ 0.0 $ (fixed) & $ 1.027 \pm 0.002 $  & (299, 293)\\
60101063002 & $ 11.01 \pm 0.60 $  & $ 1.802 \pm 0.028 $  & $ 0.060_{-0.004}^{+0.005} $  & $ -11.687_{-0.028}^{+0.026} $  & $ 0.0 $ (fixed)  & $ 1.018 \pm 0.006 $  & (345, 293)\\
60466005002 & $ 11.35 \pm 0.45 $  & $ 1.839 \pm 0.022 $  & $ 0.139 \pm 0.008 $  & $ -11.677_{-0.044}^{+0.040} $  & $ 0.0 $ (fixed)  & $ 1.003 \pm 0.005 $  & (266, 293)\\
10502008002 & $ 10.62 \pm 0.44 $  & $ 1.806 \pm 0.021 $  & $ 0.112 \pm 0.006 $  & $ -11.642_{-0.034}^{+0.031} $  & $ 0.0 $ (fixed)   & $ 1.021 \pm 0.005 $  & (305, 293)\\
\hline
\end{tabular}}\label{tb:fe_flux_fit_nustar}
  \begin{tabnote}
\footnotemark[$*$] The uncertainties in the table represent the 90\% confidence intervals.\\ 
\footnotemark[$\dagger$]Hydrogen column density along the line of sight in units of  $\rm
10^{22}\ \mathrm{cm}^{-2}$. \\
\footnotemark[$\ddagger$]The photon index of the power-law component. \\
\footnotemark[$\S$]The normalization of the power-law component at 1 keV in units of $\mathrm{photons~cm^{-2}~s^{-1}~keV^{-1}}$.\\ 
\footnotemark[$\|$] Logarithm of the Fe~K$\alpha$ line flux in units of $\mathrm{erg~cm^{-2}~s^{-1}}$ with base 10.\\
\footnotemark[$\#$]The standard deviation of the Gaussian line in units of $\mathrm{keV}$ \\
\footnotemark[$**$]Cross-normalization factors between FPMA and FPMB. \\
\end{tabnote}
\end{table*} 

\begin{table*}
\tbl{Best-fit parameters for Suzaku spectral analysis with power-law and Gaussian model\footnotemark[$*$].}{%
\begin{tabular}{lccccccc}
\hline
obsID &$N_{\mathrm{H}}^{\mathrm{LOS}}$\footnotemark[$\dagger$]& $\Gamma$\footnotemark[$\ddagger$]& norm\footnotemark[$\S$] & $\log_{10}F_\mathrm{Fe}$\footnotemark[$\|$] & $\sigma_\mathrm{Fe}$\footnotemark[$\#$] & C$_{\mathrm{BI}}$\footnotemark[$**$] &($\chi^2$, dof)\\
\hline
100005010 & $ 11.62 \pm 0.23 $  & $ 1.749 \pm 0.014 $  & $ 0.118 \pm 0.004 $  & $ -11.575_{-0.015}^{+0.014} $  & $ 0.028_{-0.008}^{+0.006} $  & $ 0.985 \pm 0.004 $  & (2690, 2586)\\
704018010 & $ 12.25 \pm 0.24 $  & $ 1.862 \pm 0.016 $  & $ 0.228 \pm 0.008 $  & $ -11.461_{-0.019}^{+0.018} $  & $ 0.015_{-0.015}^{+0.012} $  & $ 0.989 \pm 0.003 $  & (2708, 2609)\\
704018020 & $ 12.60 \pm 0.28 $  & $ 1.872 \pm 0.019 $  & $ 0.218_{-0.009}^{+0.010} $  & $ -11.427 \pm 0.019 $  & $ 0.024_{-0.019}^{+0.011} $  & $ 0.975 \pm 0.004 $  & (2439, 2444)\\
704018030 & $ 12.57 \pm 0.28 $  & $ 1.906 \pm 0.019 $  & $ 0.243_{-0.010}^{+0.011} $  & $ -11.442_{-0.022}^{+0.021} $  & $ 0.035_{-0.012}^{+0.010} $  & $ 0.972 \pm 0.004 $  & (2340, 2427)\\
708036010 & $ 10.83 \pm 0.44 $  & $ 1.809 \pm 0.029 $  & $ 0.233_{-0.015}^{+0.016} $  & $ -11.490_{-0.040}^{+0.043} $  & $ 0.015_{-0.015}^{+0.030} $  & $ 0.896 \pm 0.006 $  & (1664, 1613)\\
708036020 & $ 12.07_{-0.75}^{+0.70} $  & $ 1.867_{-0.049}^{+0.047} $  & $ 0.142_{-0.015}^{+0.016} $  & $ -11.489_{-0.032}^{+0.040} $  & $ 0.002_{-0.002}^{+0.038} $  & $ 0.888_{-0.009}^{+0.010} $ & (807, 732)\\
\hline
\end{tabular}}\label{tb:fe_flux_fit_suzaku}
  \begin{tabnote}
\footnotemark[$*$] The uncertainties in the table represent the 90\% confidence intervals.\\ 
\footnotemark[$\dagger$]Hydrogen column density along the line of sight in units of  $\rm
10^{22}\ \mathrm{cm}^{-2}$. \\
\footnotemark[$\ddagger$]The photon index of the power-law component. \\
\footnotemark[$\S$]The normalization of the power-law component at 1 keV in units of $\mathrm{photons~cm^{-2}~s^{-1}~keV^{-1}}$.\\ 
\footnotemark[$\|$] Logarithm of the Fe~K$\alpha$ line flux in units of $\mathrm{erg~cm^{-2}~s^{-1}}$ with base 10.\\
\footnotemark[$\#$]The standard deviation of the Gaussian line in units of $\mathrm{keV}$ \\
\footnotemark[$**$]Cross-normalization factors between XIS-FI and XIS-BI. \\
\end{tabnote}
\end{table*}

\begin{table*}
\tbl{Best-fit parameters for XMM-Newton spectral analysis with power-law and Gaussian model\footnotemark[$*$].}{%
\begin{tabular}{lccccccc}
\hline
obsID &$N_{\mathrm{H}}^{\mathrm{LOS}}$\footnotemark[$\dagger$]& $\Gamma$\footnotemark[$\ddagger$]& norm\footnotemark[$\S$] & $\log_{10}F_\mathrm{Fe}$\footnotemark[$\|$] & $\sigma_\mathrm{Fe}$\footnotemark[$\#$] & ($\chi^2$, dof)\\
\hline
0724060501 & $ 9.57_{-0.58}^{+0.57} $  & $ 1.603_{-0.037}^{+0.036} $  & $ 0.170 \pm 0.014 $  & $ -11.777_{-0.111}^{+0.088} $  & $ 0.005_{-0.005}^{+0.026} $ & (750, 789)\\
0724060601 & $ 10.16_{-0.60}^{+0.45} $  & $ 1.652_{-0.035}^{+0.038} $  & $ 0.171_{-0.015}^{+0.012} $  & $ -11.659_{-0.079}^{+0.070} $  & $ 0.008_{-0.007}^{+0.035} $ & (785, 763)\\
0724060701 & $ 10.22 \pm 0.62 $  & $ 1.578 \pm 0.038 $  & $ 0.068 \pm 0.006 $  & $ -11.577_{-0.036}^{+0.034} $  & $ 0.029_{-0.025}^{+0.017} $ & (939, 802)\\
0724060801 & $ 10.68 \pm 0.63 $  & $ 1.633 \pm 0.039 $  & $ 0.098 \pm 0.009 $  & $ -11.614_{-0.054}^{+0.051} $  & $ 0.041_{-0.037}^{+0.025} $ & (825, 796)\\
\hline
\end{tabular}}\label{tb:fe_flux_fit_xmm}
  \begin{tabnote}
\footnotemark[$*$] The uncertainties in the table represent the 90\% confidence intervals.\\ 
\footnotemark[$\dagger$]Hydrogen column density along the line of sight in units of  $\rm
10^{22}\ \mathrm{cm}^{-2}$. \\
\footnotemark[$\ddagger$]The photon index of the power-law component. \\
\footnotemark[$\S$]The normalization of the power-law component at 1 keV in units of $\mathrm{photons~cm^{-2}~s^{-1}~keV^{-1}}$.\\ 
\footnotemark[$\|$] Logarithm of the Fe~K$\alpha$ line flux in units of $\mathrm{erg~cm^{-2}~s^{-1}}$ with base 10.\\
\footnotemark[$\#$]The standard deviation of the Gaussian line in units of $\mathrm{keV}$
\end{tabnote}
\end{table*}

\begin{table*}
\tbl{Best-fit parameters for Swift/XRT spectral analysis with power-law and Gaussian model\footnotemark[$*$].}{%
\begin{tabular}{lccccc}
\hline
obsID &$N_{\mathrm{H}}^{\mathrm{LOS}}$\footnotemark[$\dagger$]& $\Gamma$\footnotemark[$\ddagger$] & $\log_{10}F_\mathrm{Fe}$\footnotemark[$\S$] & (Cstat, dof)\\
\hline
00031312009--00031312038 & $ 11.56 \pm 1.03 $  & $ 1.695_{-0.077}^{+0.078} $   & $ -11.592_{-0.110}^{+0.089} $ & (13804, 13941)\\
00031312050--00031312094 & $ 14.98_{-0.76}^{+0.77} $  & $ 2.011 \pm 0.058 $  & $ -11.619_{-0.083}^{+0.070} $ & (22340, 21909)\\
\hline
\end{tabular}}\label{tb:fe_flux_fit_swftxrt}
  \begin{tabnote}
\footnotemark[$*$] The uncertainties in the table represent the 90\% confidence intervals. The normalization of the power-law component of each data was omitted.\\ 
\footnotemark[$\dagger$]Hydrogen column density along the line of sight in units of  $\rm
10^{22}\ \mathrm{cm}^{-2}$. \\
\footnotemark[$\ddagger$]The photon index of the power-law component. \\
\footnotemark[$\S$] Logarithm of the Fe~K$\alpha$ line flux in units of $\mathrm{erg~cm^{-2}~s^{-1}}$ with base 10.\\
\end{tabnote}
\end{table*}

\begin{figure*}
 \begin{center}
  \includegraphics[width=12cm]{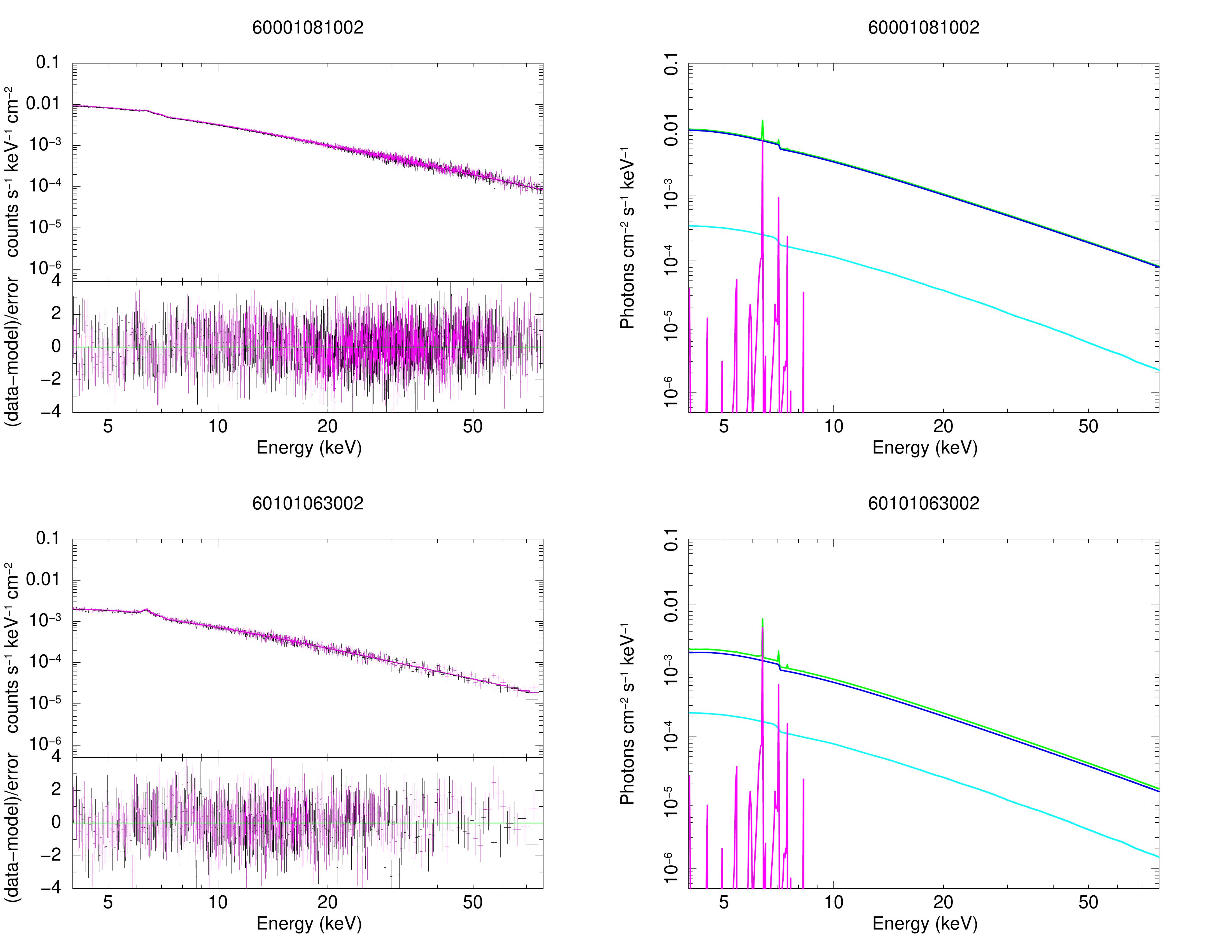}
 \end{center}
 \caption{Left: folded X-ray spectra fitted with XClumpy model. The black and magenta crosses are NuSTAR/FPMA and FPMB data, respectively. The solid curves represent the best-fit model. The lower panel shows residuals. Right: best-fit model components for FPMA. Green lines are total, blue lines are direct components, light blue lines are reflection continuum from the torus, and magenta lines are emitted lines from the torus.}
 \label{fig:spec_nustar1}
\end{figure*}

\begin{figure*}
 \begin{center}
 \includegraphics[width=12cm]{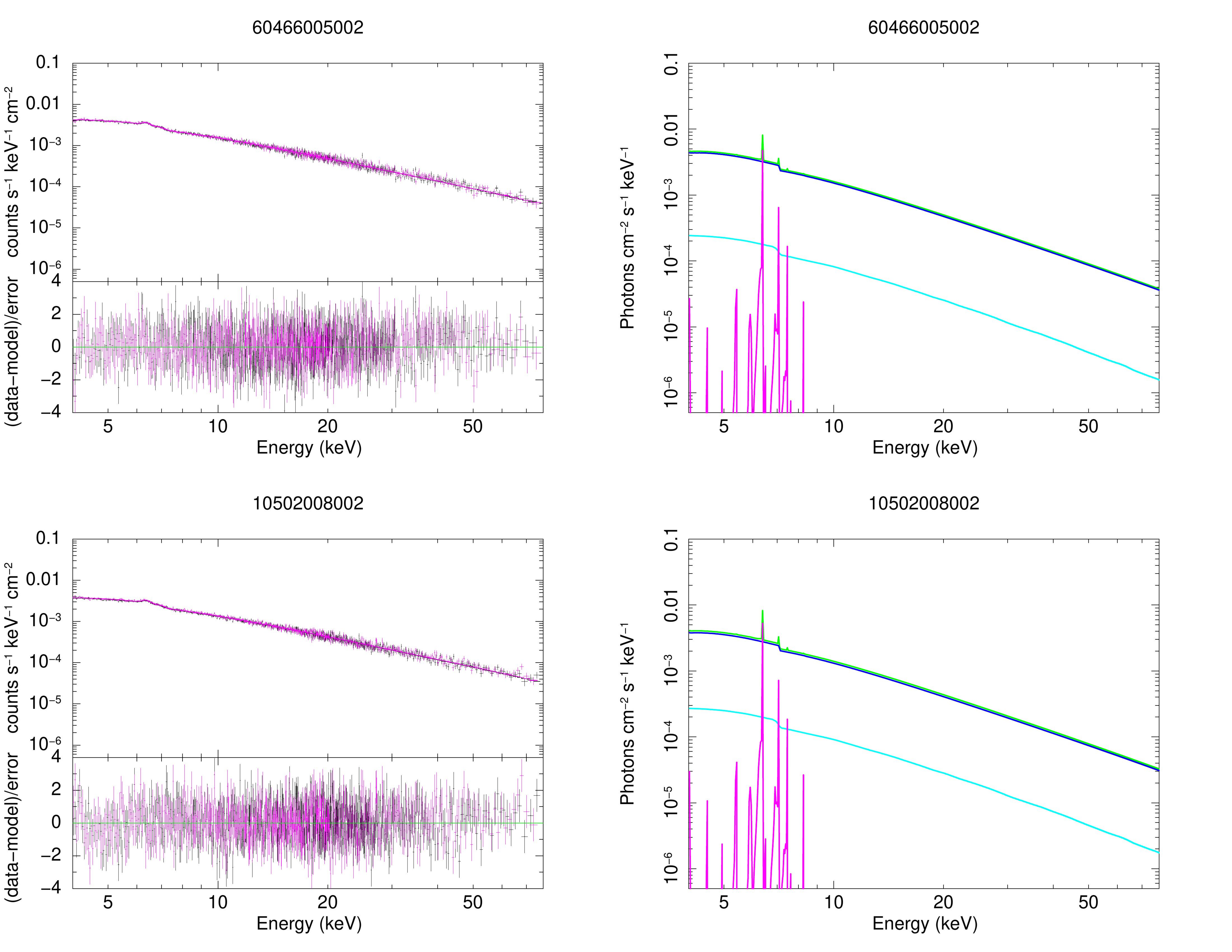}
 \end{center}
 \caption{Continued.}
 \label{fig:spec_nustar2}
\end{figure*}

\begin{figure*}
 \begin{center}
 \includegraphics[width=12cm]{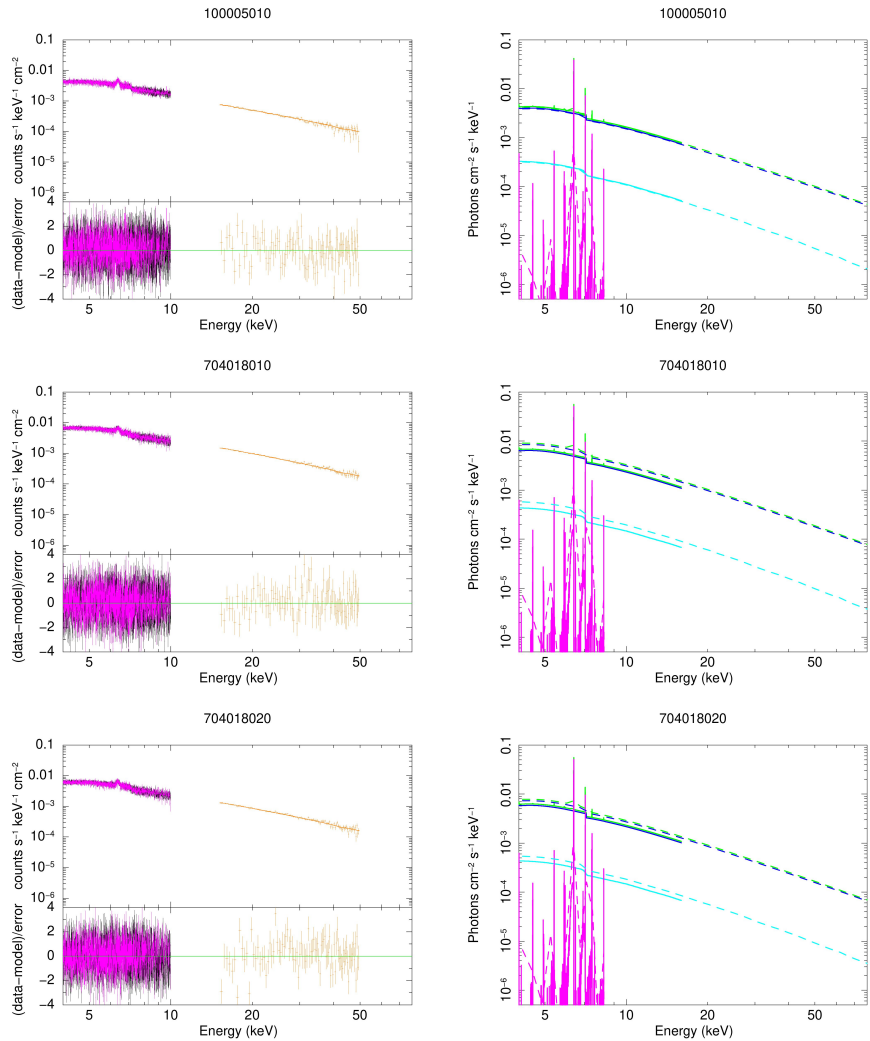}
 \end{center}
 \caption{Left: folded X-ray spectra fitted with XClumpy model. The black, magenta, and orange crosses are Suzaku/XIS-FI, XIS-BI, and HXD-PIN data, respectively. The solid curves represent the best-fit model. The lower panel shows residuals. Right: best-fit model components for Suzaku/XIS-FI (solid lines) and HXD-PIN (dashed lines). Green lines are total, blue lines are direct components, light blue lines are reflection continuum from the torus, and magenta lines are emitted lines from the torus.}
 \label{fig:spec_suzaku1}
\end{figure*}

\begin{figure*}
 \begin{center}
 \includegraphics[width=12cm]{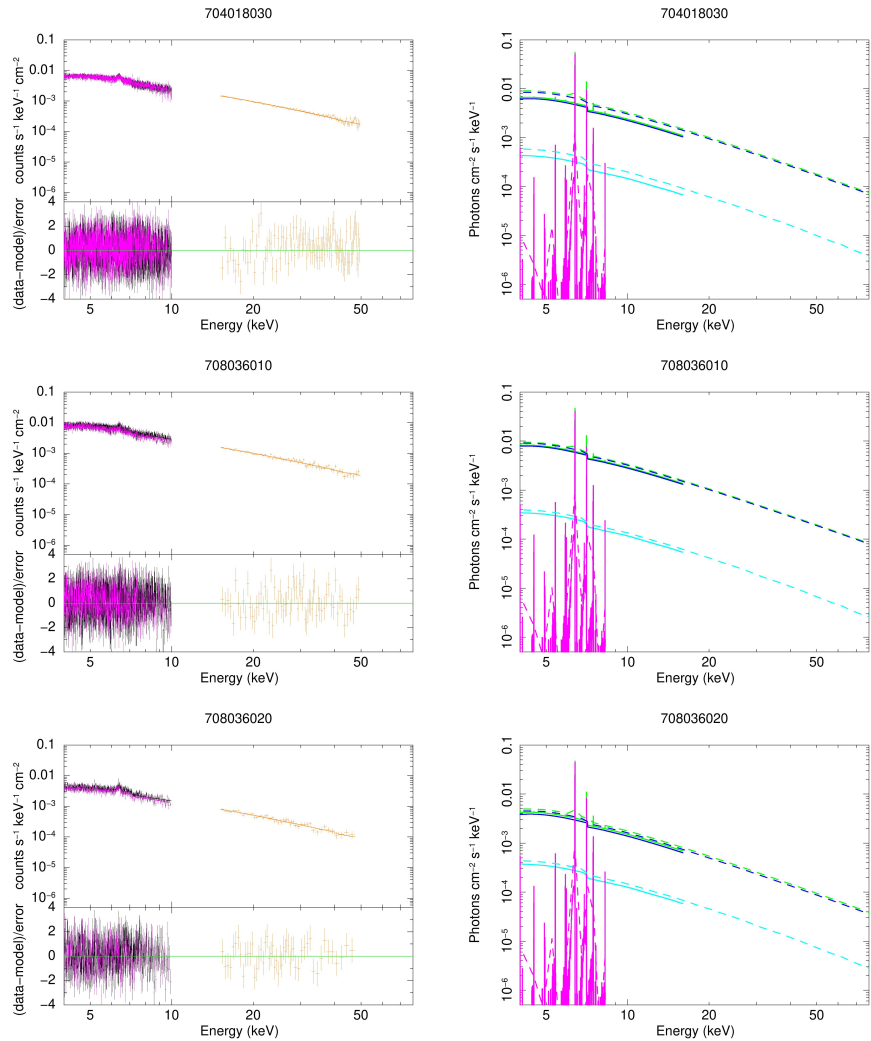}
 \end{center}
 \caption{Continued.}
 \label{fig:spec_suzaku2}
\end{figure*}

\bibliography{cenA.bib}

\end{document}